\documentclass[aps,showpacs,nofootinbib, epsf,twocolumn,pra]{revtex4}
\usepackage{amsmath}
\usepackage{amssymb}
\usepackage{bbold}
\usepackage{mathrsfs}
\usepackage{graphicx, psfrag}
\usepackage[centerlast]{caption}
\usepackage{float}
\usepackage{subfig}
\usepackage{tikz}
\usepackage{pgfplots}
\usepackage{braket} 
\usepackage[colorlinks=true, citecolor=blue, urlcolor = blue, linkcolor= red,bookmarks=true]{hyperref}
\usepackage{epstopdf}
\graphicspath{{figures/}}
\newenvironment{rcases}
  {\left.\begin{aligned}}
  {\end{aligned}\right\rbrace}
\DeclareCaptionLabelFormat{cont}{#1~#2\alph{ContinuedFloat}}
\usepackage[normalem]{ulem}
\def \beq{\begin{equation}}
\def \eeq{\end{equation}}
\def \bse{\begin{subequations}}
\def \ese{\end{subequations}}
\def \bea{\begin{eqnarray}}
\def \eea{\end{eqnarray}}
\def \bem{\begin{displaymath}}
\def \eem{\end{displaymath}}
\def \bem{\begin{bmatrix}}
\def \eem{\end{bmatrix}}
\def \bs{\boldsymbol} 
\def \bb{\bibitem}
\def \nn{\nonumber}
\def \bc{\begin{center}}
\def \ec{\end{center}}
\begin{document}
\title{\textbf{Dimensional crossover in self-organised super-radiant phases of ultra cold atoms inside a cavity}}
\author{Poornima Shakya, Amulya Ratnakar\footnote{Current Address : Indian Institute of Science Education And Research Kolkata, Mohanpur, Nadia - 741 246 West Bengal, India}, and Sankalpa Ghosh}
\affiliation{Department of Physics, Indian Institute of Technology Delhi, New Delhi-110016, India}

\begin{abstract}
We consider a condensate of ultra cold bosonic atoms in a linear optical cavity illuminated by a two-pump configuration where each pump is making different angles with the direction of the cavity axis. We show such configuration allows a smooth transition from a one-dimensional quantum optical lattice configuration to a two-dimensional quantum optical lattice configuration induced by the cavity-atom interaction.
Using a Holstein-Primakoff transformation, we find out the atomic density profile of such self-organised ground state in the super-radiant phase as a function of the angular orientations of the pump in such dynamical quantum optical lattice, and, also provide an analysis of their structures in coordinate 
and momentum space. 
In the later part of the paper, we show how the corresponding results can also be qualitatively understood in terms of an Extended Bose-Hubbard model in such quantum optical lattice potential.  
\end{abstract}
\maketitle
\section{Introduction}
The pioneering work of Dicke \cite{Dicke} that predicts the excitation of a super-radiant phase \cite{Lieb, Wang, Haroche} by a radiation pulse obtained its convincing experimental demonstration in the system of ultra cold atoms inside a cavity by the Esslinger group \cite{Baumann}, which observed  a normal to super-radiant phase transition in ultra cold atomic Bose-Einstein condensate
through self-organisation \cite{Nagy}. This observation was preceded by the experimental observation of super-radiant Rayleigh scattering from ultra cold atoms in free-space \cite{Rayleigh} and ring cavities \cite{Ring}. 
It opened up a new direction in the study of exotic quantum many-body phases \cite{Sarang1, Sarang2, RitschRev, RitschRev2} with ultra cold atomic condensate trapped in optical lattice potentials. 

Initial study of such quantum many-body phases of ultra-cold atomic systems involve free space optical lattice potentials that 
are not affected by the atomic density distribution \cite{Zoller, Greiner}
and hence the optical-lattice potential acts as a classical external potential on the ultra cold atoms \cite{Lewenstein,Bloch} and does not have their dynamics. 
As compared to that, atomic condensates trapped in cavity-generated dynamical quantum optical lattice potential \cite{Domokos, Asboth, Larson, Larson1} have a significant impact on the structure and strength of the lattice potential \cite{RitschRev}. 
Particularly, when  these trapped atoms are directly illuminated by a transverse pump beam, then the excited atoms scatter the pump photons, which finally populate the cavity mode. The position-dependent atom-photon coupling gives rise to a position-dependent scattered-field amplitude. It can generate novel self-organised quantum-many body phases of the atoms through the cavity-mediated long-range interactions \cite{Maschler} such as a lattice super-solid phase through Dicke type of transition \cite{Baumann}, or a more ideal super-solid phase \cite{Leonard, Gremaud, Zwerger, Morales} where a continuous gauge symmetry and a continuous translational symmetry is spontaneously 
broken \cite{Guo} leading to the simultaneous existence of an off-diagonal long-range order (a property of superfluid) and a diagonal long-range order (property of a solid) \cite{ Andreev, Legett, Matsuda, Liu, Nelson, Bruder, Roddick, Otterlo, Batrouni, Scalettar, Boninsegni, Meisel, Greywall, Bishop, Goodkind, Chan, Kim, Beamish, Chan1}. 

The existence of competing short-range and cavity-induced long-range interaction in the bosonic lattice model provides a host of novel quantum phases, such as superfluid, supersolid, Mott insulator, and charge density wave \cite{Landig, Huang}, their novel collective excitations \cite{Dogra}, metastability and avalanche dynamics in Mott-insulator and density-wave phases \cite{Hruby}. The transition from a coherent superfluid phase to an incoherent Mott-insulator, both lying in a super-radiant regime, was also studied by combining Bose-Hubbard Model with the Dicke model \cite{Hemmerich} and also by using multi-configuration time-dependent Hartree method for 
indistinguishable particles\cite{MCTDH}.
The quantum properties of light also get significantly modified due to the interplay of the cavity-mediated long-range interactions and the short-range processes of the atom \cite{Benitez1}. They can be designed and optimised to create new types of quantum simulators \cite{Benitez2} both for a single and a multi-mode cavity. 
Other significant works in this direction explored quantum magnetism by simulating quantum spin Hamiltonian with multi-component ultra cold atoms in a linear cavity pumped by external lasers \cite{Mivehvar1}, p-band induced self-organisation and dynamics in an optical cavity \cite{Zupancic},
spin-entanglement and magnetic competition in spinor quantum optical lattices in a cavity \cite{Benitez3}, creation of various topologically non-trivial phases in a cavity atom system \cite{Mivehvar2, Morigi1}, parametric instabilities in a driven-dissipative Bose-Einstein condensate (BEC) in a cavity \cite{Chitra1}, 
dissipation engineered family of dark states in cavity-atom systems \cite{Chitra2}, role of the atomic correlations in the dynamical instability generated in a cavity-BEC system \cite{Chitra3}, possibility of the existence of intertwined and vestigial order in a crossed-cavity-multimode BEC system \cite{Demler}, super-radiant scattering and dynamical instability in a system of a linear cavity illuminated by a single pump\cite{Piazza_ritsch}, recent observation of a time crystal stabilised by dissipation in a driven open cavity-BEC system \cite{Kebler} etc. Cavity-like periodic patterns in the atomic density have also been observed in free-space systems under certain conditions \cite{Ostermann} and 
Dicke superradiance was also  studied in fermionic gases \cite{Keeling}.
  
Most of the works mentioned above considered single pump laser in a linear cavity 
\cite{Baumann, Sarang1, Domokos, Asboth, Larson, Larson1, Maschler, Landig, Huang, Dogra, Hruby, Hemmerich, MCTDH, Benitez1, Benitez2, Mivehvar1, Zupancic, Benitez3, Mivehvar2, Morigi1, Chitra1, Chitra2, Chitra3, Piazza_ritsch, Kebler, Ostermann, Keeling, Nagy},
 ring cavity \cite{Ring}, or crossed cavity \cite{Leonard, Gremaud, Zwerger, Demler} set-up. This does not change the dimensionality of the self-organised super-radiant phases and their corresponding quantum optical lattice potential for a given set-up.
It may be noted that in the simulation of various quantum, many-body phases that have a breadth-taking spread from hard condensed matter problems to the systems studied in high energy physics ( for a review, see, e.g. \cite{Gross}) 
with table-top ultra cold atomic systems in a classical optical lattice, the dimensionality of the classical optical lattice plays a significant role \cite{Blochreview} in determining the nature of the simulated quantum system. The variety of the quantum systems that can be simulated by ultra cold atoms can be significantly enhanced if the dynamical quantum optical lattice potential created inside an optical cavity can also 
be made into different dimensions and that too in an interchangeable way. By considering an ultra cold Bose-Einstein condensate (BEC) placed inside a linear cavity illuminated by two pump beams making angles $\theta_1$ and $\theta_2$ with the cavity axis, in the current work, 
we proposed a simple way of achieving this by varying the relative angle between these classical pumps. 
 
The existence of two tuning parameters $\theta_1$ and $\theta_2$ offers the possibility of realisation of a large number of 
self-organised phases. We demonstrate that by changing these angles in the super-radiant regime, one can continuously transform from a one-dimensional self-organised (SO) lattice-supersolid phase to a two-dimensional SO lattice supersolid phase. 
The dimensionality was clearly identified by identifying the atomic density maxima and minima in coordinate space and analysing them in the momentum space. These findings form one of the main results of our paper.
Using a Holstein-Primakoff (HP) transformation \cite{Emary1, Emary}, we determine the dynamically generated quantum optical lattice potential in these SO phases, corresponding atomic density distribution that shows the change in dimensionality as the angles made by two pumps with the cavity axis are varied in the range $(\theta_1,\theta_2)\in[0,\pi/2]$. We additionally show how such SO phases evolve as a function of the increasing intra-cavity photon number. Our proposed set-up enjoys the possibility of experimental realisation since this is an intermediate between the already realised experimental set-up of a single cavity illuminated by a  single transverse pump  \cite{Baumann}  and the crossed-cavity \cite{Gremaud, Zwerger} illuminated by a single pump \cite{Leonard, Morales}. It can also be generalised to other cavity set-ups in addition to the linear one we have considered here.

To consolidate our theoretical analysis further, in the later part of this work using the tight-binding approximation in the dynamical quantum optical lattice potential,
we derive an extended Bose-Hubbard model (EBHM) model Hamiltonian for our system (for a detailed review of various Hubbard models in ultra cold atomic systems, see \cite{Hubbardreview}) in certain ranges of the cavity parameters in terms of the atomic-field operators only
from the microscopic many-body Hamiltonian for such atom-photon system. 
The EBHM written in this form makes it easier to capture the cavity-mediated long-range interaction, which is responsible for the various self-organized lattice supersolid phases that we observed. We pointed out when such EBHM is relatively more useful as compared to HP approximation to describe the super-radiant phase inside a cavity.
Consequently, this allows us to compare our 
approach to study such SO phases  with the  
EBHM derived for ultra cold atomic systems' classical optical lattice potential \cite{Kuhner, Schmid, Kovrizhin, Pai} that also predicts a transition from density wave to supersolid phase. 
We also evaluate the quantum optical lattice potential using this EBHM and compare it with the same obtained under HP transformation and the balanced pump condition ( to be defined later).

Accordingly, we organise the rest of the paper in the following way. In Section \ref{MS}, we introduce the model microscopic Hamiltonian for our system and discuss the scattering states of the atoms. 
In Section \ref{HPA}, we introduce the HP approximation to calculate the properties of the super-radiant phases and show the method of calculation within this approach in detail. In Section \ref{RD_sec}, we discuss the main results of this paper under HP approximation. In \ref{sec_QOL},
we provide the results for the dynamical optical lattices and in \ref{sec_dens}, we provide the results for the ground state atomic density in the self-organised phases. In Section \ref{class}, we classify the self-organised phases, and identify their dimensionality by evaluating the maxima and minima points of the atomic density.
We also provide an analysis of the momentum spectrum of such self-organised phases in Section \ref{sec_TOF} that can be related with the time of flight images of the condensate in the related experiments. 
In Section \ref{BHM}, we construct the EBHM for this system and discuss its connection with the HP formalism in the earlier section. Finally, in the conclusion section, we summarise the significance of our analysis. Appendices contain some details that are linked to various portions of the main text. 

\begin{figure*}
\includegraphics[width=2\columnwidth, height=1\columnwidth]{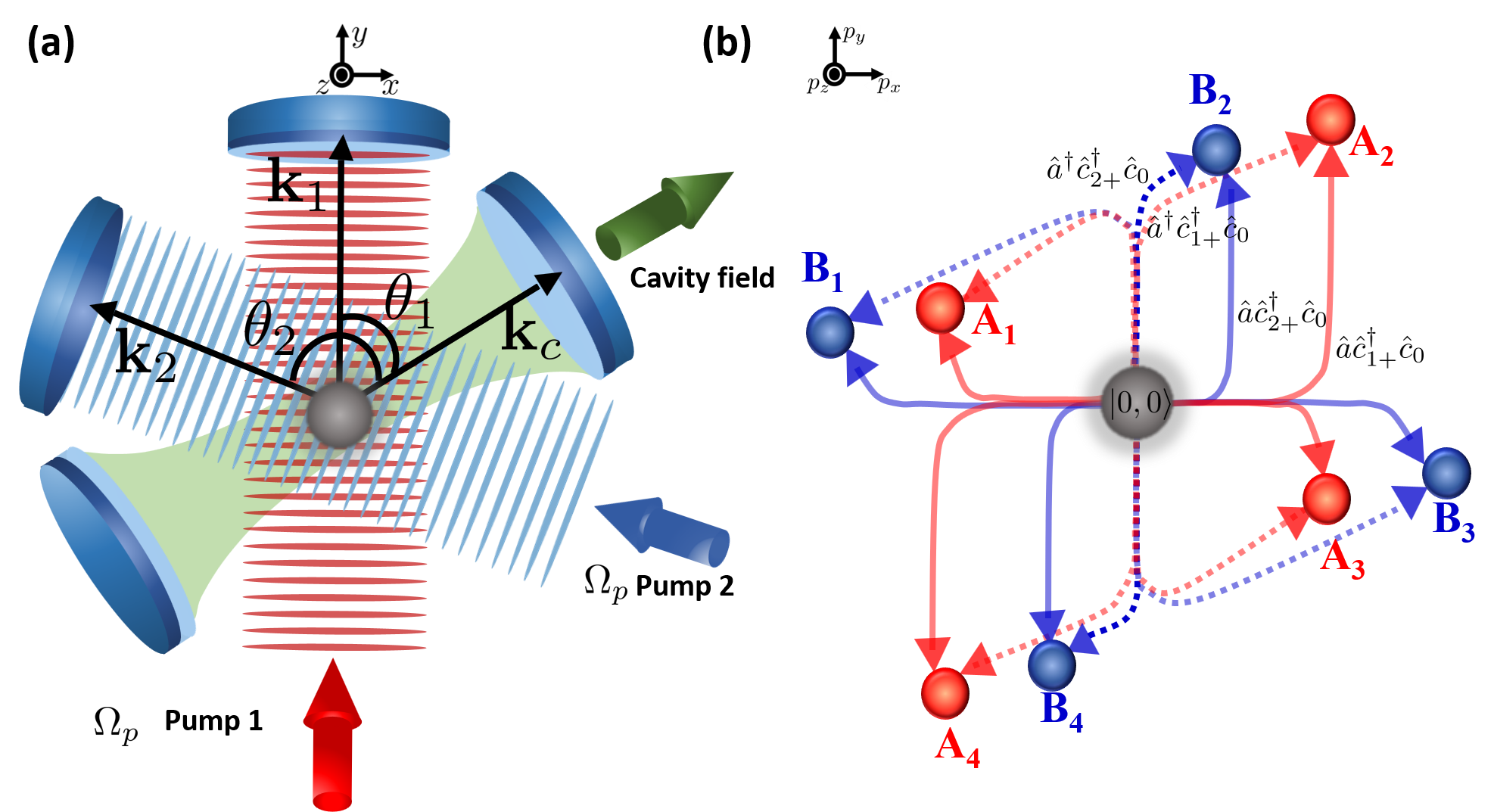}
\caption{(\textit{color online}) \small (a) Schematic for the single cavity - two pump system. The atomic condensate is trapped at the intersection of the two pump beams. Pump 1 makes an angle $\theta_1$ and pump 2 makes an angle $\theta_2$ with the cavity axis. Wave-vectors $\bs{k}_{1,2,c}$ and $\Omega_{p}$ are explained in the text.(b) The momentum diagram shows the 9 momentum states in terms of $\theta_1$ and $\theta_2$ - $\vert 0,0\rangle$, $A_1=\vert -\hbar k \sin\theta_1, \hbar k (1-\cos\theta_1)\rangle$,  $A_2=\vert\hbar k \sin\theta_1, \hbar k (1+\cos\theta_1)\rangle$,  $A_3=\vert \hbar k \sin\theta_1, \hbar k (-1+\cos\theta_1)\rangle$,  $A_4=\vert -\hbar k \sin\theta_1, \hbar k (-1-\cos\theta_1)\rangle$,  $B_1 =\vert \hbar k (-\sin\theta_1-\sin(\theta_2-\theta_1)), \hbar k (-\cos\theta_1+\cos(\theta_2-\theta_1))\rangle$,  $B_2=\vert\hbar k (\sin\theta_1-\sin(\theta_2-\theta_1)), \hbar k (\cos\theta_1+\cos(\theta_2-\theta_1))\rangle$,  $B_3=\vert \hbar k (\sin\theta_1+\sin(\theta_2-\theta_1)), \hbar k (\cos\theta_1-\cos(\theta_2-\theta_1))\rangle$ and $B_4=\vert \hbar k (-\sin\theta_1+\sin(\theta_2-\theta_1)), \hbar k (-\cos\theta_1-\cos(\theta_2-\theta_1))\rangle$. The annotations of the 
color, dotted and dashed lines is explained in the text in Section(\ref{MS})}\label{schematic}
\end{figure*}
\section{Model system and the Hamiltonian}\label{MS}
We consider a linear cavity with a single cavity mode characterised by frequency $\omega_{c}$ and wave-vector $\bs{k}_{c}$, illuminated by two pump beams at frequency $\omega_p$ making an angle, $\theta_1$ and $\theta_2$, with the cavity axis. The cavity is loaded with a Bose-Einstein condensate, with $N=1.05\times 10^5$ ${^{87}}{Rb}$ atoms in the $\vert F,m_F\rangle = \vert 1,-1\rangle$ state, where $F$ and $m_F$ are the total angular momentum and the corresponding magnetic quantum number. The cavity is detuned from the pump laser frequency by $\Delta_c = \omega_p-\omega_c$. $\hat{a}(\hat{a}^{\dagger})$ is the annihilation (creation ) operator which annihilates (creates) a photon in the cavity mode with wave vector, $\bs{k_c}$.
 We have taken pump 1 to be along the $y-$ direction and this choice gives 
\bea \bs{k}_{c}  & = & k \sin(\theta_1) \hat{x} + k \cos (\theta_1) \hat{y}. \label{kc} \\
 \bs{k}_{1}  &= &  k\hat{y},~\bs{k}_{2}  = -k \sin(\theta_2-\theta_1) \hat{x} + k \cos (\theta_2-\theta_1) \hat{y} \label{k12} \eea 
 where  $k =\frac{2\pi}{\lambda_p}$ with  $\lambda_p$ being the pump wavelength. 
\begin{figure*}
	\includegraphics[width=2.\columnwidth, height=1.4\columnwidth]{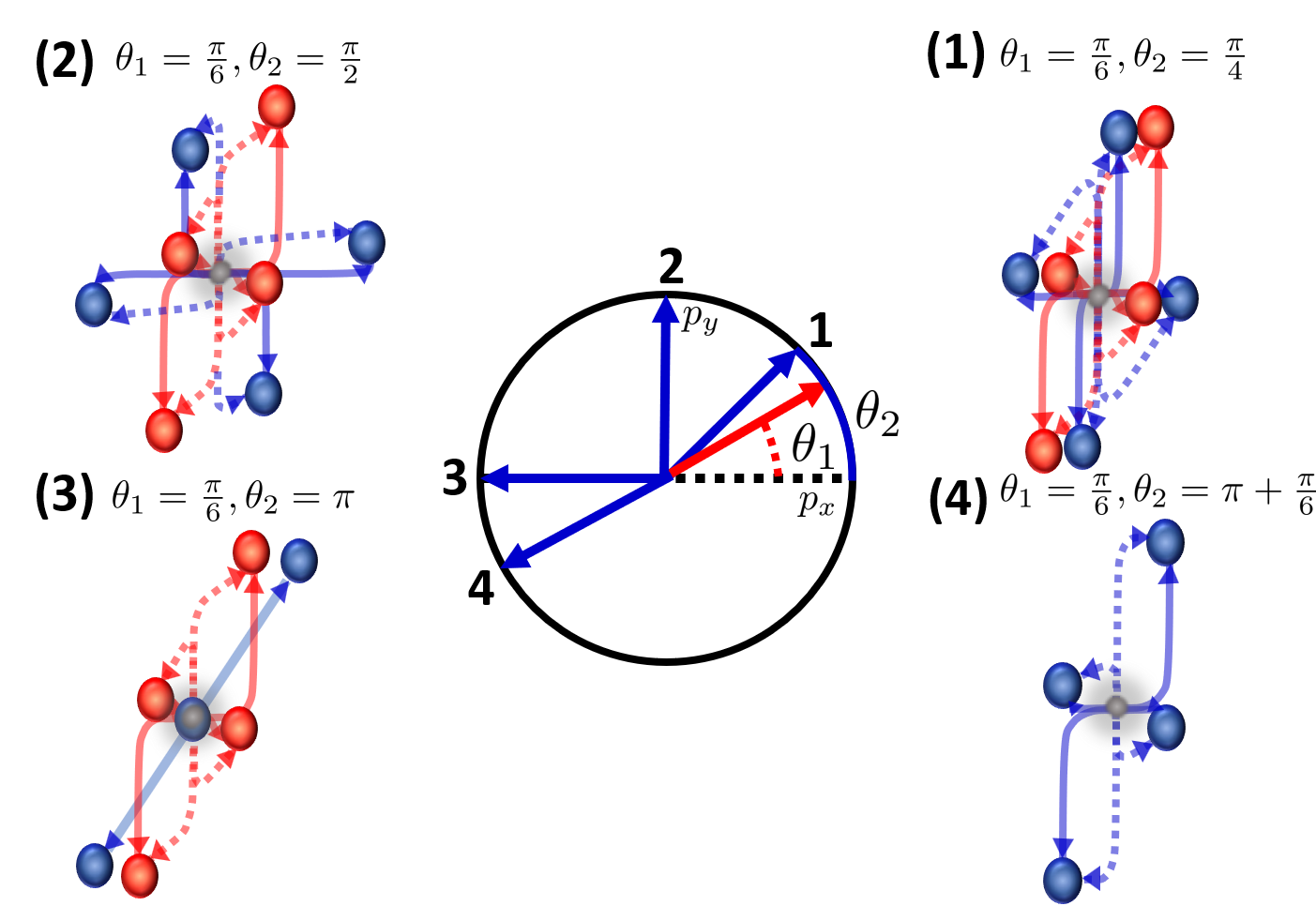}
\caption{(\textit{color online}) We provide the momentum diagrams for the two pump-cavity system for few specific values of the two pump angles $\theta_1 $ and $\theta_2$ which are indicated in each figure. In (3), three momentum states with same $\ket{k_{x}, k_{y}}$ coincide at the origin where as in (4) all the four {\it blue} and {\it red} momentum states coincide. The annotations of the color, dotted and dashed lines are same as in Fig. \ref{schematic}(b).}\label{mom_2pump}
\end{figure*}

The microscopic many-body Hamiltonian for the system is written as-
\begin{widetext}
\beq
\hat{H}_{MB} = \left(-\hbar\Delta_{c}\hat{a}^{\dagger}\hat{a}\right)+\int\int_A \hat{\Psi}^{\dag}(x,y)\left(\frac{\hat{p}^{2}_{x}+\hat{p}^{2}_{y}}{2M_a} 
+ \hat{V}(\bs{r}) + \frac{g_{2D}}{2}\vert\hat{\Psi}(x,y)\vert^{2}-\mu_0\right)\hat{\Psi}(x,y)dxdy \label{2pump}
\eeq
where 
\bea
\hat{V}(\bs{r}) & = & \hbar \eta \cos(\bs{k}_{1}\cdot\bs{r} ) \cos(\bs{k}_{c}\cdot\bs{r})(\hat{a}^{\dagger} + \hat{a}) +\hbar \eta \cos(\bs{k}_{2}\cdot\bs{r} ) \cos(\bs{k}_{c}\cdot\bs{r})(\hat{a}^{\dagger} + \hat{a}) + \hbar U_{0} \cos^{2}(\bs{k}_{c}\cdot\bs{r})\hat{a}^{\dagger}\hat{a}\nn\\
&+& \hbar U_{p}\cos^{2}(\bs{k}_{1}\cdot\bs{r}) +\hbar U_{p}\cos^{2}(\bs{k}_{2}\cdot\bs{r}) + \hbar U_{p}\cos(\bs{k}_{1}\cdot\bs{r} ) \cos(\bs{k}_{2}\cdot\bs{r}) ,\label{pot_pump}
\eea
\end{widetext}
and $\hat{\Psi}(x,y)$ ($\Psi^{\dag}(x,y)$) is the atomic field operator which annihilates(creates) a particle at position $(x,y)$. The number operator is $\hat{N}=\int d\vec{r} \hat{\Psi}^{\dag}(\vec{r})\hat{\Psi}(\vec{r})$. $U_p = \Omega_p^2/\Delta_a$ is the potential depth of the transverse pump potentials formed by the two pumps where $\Omega_{p}$ is the maximum pump Rabi frequency. $g_{0}$ is the maximum atom-photon coupling strength and $U_0 = g^2_{0}/\Delta_a$ is the depth of the potential formed by the cavity field. $U_0$ denotes the maximum shift in the resonance frequency for a single intracavity photon. $\eta = \Omega_{p}g_{0}/\Delta_{a}$ is the two-photon Rabi frequency for the cavity and signifies the strength of the interaction between the pumps and the cavity field.  $g_{2D} = 4\pi a_s \hbar^2 AN/m_a V$ is the strength of the short-range $s$-wave collisions with scattering length $a_s$. $V = 4\pi r_x r_y r_z/3$, $A$ is the area of the unit cell formed by the interference pattern between the cavity and the two pump modes, 
 and $r_x$, $r_y$ and $r_z$ are the Thomas-Fermi radii along the $x, y$ and $z$ directions. $\mu_0$ is the chemical potential.

The schematic for the system is given in Fig.\ref{schematic}(a).
The pump modes are far red-detuned $(\Delta_a=\omega_p-\omega_a \ll 0,~\vert \Delta_a\vert>>g_{0}, \Delta_{c})$ from the atomic transition frequency $\omega_a$ and hence, atomic transition of the atoms to the internal excited state is suppressed. 
This allows us to adiabatically eliminate the excited state and the atoms initially prepared in their internal ground state, mostly evolve in the ground state. This gives rise to the atom-cavity interaction  
which can be seen as a dynamical (quantum) optical lattice potential $\hat{V}(\bs{r})$ given in  Eq.(\ref{pot_pump})
with its depth depending on the cavity field amplitudes.

The cavity does not contain any photons initially. The transverse pump beams are closely detuned with the cavity resonance. The excited atoms coherently scatter the pump photons into the cavity mode via off-resonant Raman scattering processes \cite{Domokos,Maschler}. These processes couple the BEC zero momentum mode $\vert 0,0\rangle$ to the eight momentum modes, $\vert\pm \hbar \bs{k}_c \pm \hbar \bs{k}_1\rangle$ and $\vert\pm \hbar \bs{k}_c \pm \hbar \bs{k}_2\rangle$ where $\bs{k}_{c}$ and $\bs{k}_{1,2}$ are defined in Eq. (\ref{kc}) and (\ref{k12}). The explicit tabulation of these momentum states are given in the caption of Fig. \ref{schematic} (b). 
 
To gain more insight about effect of orientation of the two pumps,  we additionally provide 
 the momentum diagrams for few angles in Fig.(\ref{mom_2pump}). 
 The $red$ arrow in the central circle in Fig.(\ref{mom_2pump}) shows the angle of pump 1 ($\theta_1$) and the $blue$ arrow shows the angle of pump 2 ($\theta_2$) with the cavity axis. We keep $\theta_1$ constant and vary $\theta_2$. The details of the figure are explained in the captions. 
 We start with case (1), $\theta_1=\pi/6, \theta_2=\pi/4$ where we show the eight separate momentum states along with the zero momentum state (appears in $grey$ colour at co-ordinate $\vert k_x,k_y\rangle=\vert 0,0\rangle$) as shown in Fig. \ref{schematic}(b).
 Then as $\theta_{2}$ increases to $\frac{\pi}{2}$ in case (2), we see 
 that the $blue$ momentum states are rotating by the same amount as there is a change in the value of $\theta_2$, therefore, transferring the rotation in real space to that in the reciprocal space. 
 In case (3), $\theta_1=\pi/6, \theta_2=\pi$, two $blue$ momentum states coincide with zero-momentum state. We define this coincidence as the degeneracy in the momentum space. 
 As expected  these momentum states only depend on the angle between the two pumps, namely $(0<\theta_2-\theta_1<\pi/2)$. In subsequent discussion, using a Holstein-Primakoff (HP) transformation, we shall see the effect of these angular change on the self-organised phases inside the cavity. 

\section{Holstein-Primakoff approach}\label{HPA}
We expand the atomic field operator in the momentum modes, resulting from the scattering processes. The expression for the atomic field operator is -
\beq
\hat{\Psi}(x,y) = \psi_{0}\hat{c}_{0} + \psi_{1-}\hat{c}_{1-} + \psi_{1+}\hat{c}_{1+}+ \psi_{2-}\hat{c}_{2-} + \psi_{2+}\hat{c}_{2+} \label{atomic_op} 
\eeq
where $\psi_{0} = \sqrt{1/A}$ represents the BEC zero-momentum mode, 
$\psi_{1\pm}= \sqrt{2/A} \cos[(\bs{k}_{c} \pm \bs{k}_{1})\cdot\bs{r}]$ and  $\psi_{2\pm}= \sqrt{2/A} \cos[(\bs{k}_{c} \pm \bs{k}_{2})\cdot\bs{r}]$
represent the atomic modes with momenta $\bs{k}_c\pm\bs{k}_1$ and $\bs{k}_c\pm\bs{k}_2$ respectively.  $\hat{c}^{\dag}_{0}$, $\hat{c}^{\dag}_{1\pm}$ and $\hat{c}^{\dag}_{2\pm}$ respectively 
creates an atom at $\ket{0,0}$,  an excitation with energy $\hbar\omega_{1\pm}$ and an excitation with energy $\hbar\omega_{2\pm}$. 
We substitute the expansion of the atomic field operator in Eq.$(\ref{2pump})$ to obtain the following effective many-body Hamiltonian -
\bea 
\hat{H}_{MB}&=&  -\hbar\bar{\Delta}_c\hat{a}^{\dagger}\hat{a}+\sum_{i=\pm1,\pm2}\hbar\omega_{i} \hat{c}^{\dagger}_{i}\hat{c}_{i} \nonumber\\
&+&\frac{\hbar \lambda}{\sqrt{N}}\left(\hat{a}^{\dag}+\hat{a}\right) \sum_{i=\pm1,\pm2}\left(\hat{c}^{\dag}_{i}\hat{c}_{0}+\hat{c}^{\dag}_{0}\hat{c}_{i}\right)\label{2pump1} \eea
where $\omega_{\text{rec}}$ is the frequency associated with the recoil energy, $E_{rec}=\hbar \omega_{rec}$. $\omega_{rec}=\hbar k^2/2M_a, \omega_{1+}=2(1+\cos\theta_1)\omega_{rec}$, $\omega_{1-}=2(1-\cos\theta_1)\omega_{rec}$, $\omega_{2+}=2(1+\cos(\theta_2))\omega_{rec}$, $\omega_{2-}=2(1-\cos(\theta_2))\omega_{rec}$ and $\bar{\Delta}_c=\omega_p-\omega_c-\frac{U_0N}{2}$. $\lambda=\frac{\eta \sqrt{N}}{2\sqrt{2}}$ is the coupling parameter between the atom and the two pump modes. The atomic momentum states in Fig.(\ref{schematic}b) and Fig.(\ref{mom_2pump}) describe the scattering of a photon with momentum $\hbar k$ from pump 1 ($red$) and pump 2 ($blue$) into the cavity mode. There are two ways to reach the excited momentum states $\vert \pm \hbar \bs{k}_{c}\pm\hbar \bs{k}_{1,2}\rangle$ from the BEC zero-momentum state $\vert 0,0\rangle$. The dotted $red$ ($blue$) lines show the absorption of pump 1 (pump 2) photon accompanied by emission of a photon into the cavity and is identified by the operator $\hat{a}^{\dag}\hat{c}_{1+}^{\dag}\hat{c}_0$($\hat{a}^{\dag}\hat{c}_{2+}^{\dag}\hat{c}_0$). The solid $red$ ($blue$) lines show the absorption of a cavity photon accompanied by emission of the photon into pump 1 (pump 2) and is identified by the operator $\hat{a}\hat{c}_{1+}^{\dag}\hat{c}_0$($\hat{a}\hat{c}_{2+}^{\dag}\hat{c}_0$). The reverse processes are not shown in the diagram. These processes correspond to the operators $\hat{a}^{\dag}\hat{c}_{0}^{\dag}\hat{c}_{1+}$($\hat{a}^{\dag}\hat{c}_{0}^{\dag}\hat{c}_{2+}$) and  $\hat{a}\hat{c}_{0}^{\dag}\hat{c}_{1+}$($\hat{a}\hat{c}_{0}^{\dag}\hat{c}_{2+}$).  

To underscore the similarity of the Hamiltonian (\ref{2pump1}) with the proto-type Dicke model, we use the generalized Holstein-Primakoff transformation \cite{Emary} -
\bea
\begin{rcases}
\hat{c}^{\dag}_{p}\hat{c}_{q}&=& \hat{b}^{\dag}_{p}\hat{b}_{q},\\
\hat{c}^{\dag}_{p}\hat{c}_{0}&=& \hat{b}^{\dag}_{p}\hat{\Theta}_{0}(N),\\
\hat{c}^{\dag}_{0}\hat{c}_{q}&=& \hat{\Theta}_{0}(N)\hat{b}_{q},\\
\hat{c}^{\dag}_{0}\hat{c}_{0}&=& \hat{\Theta}_{0}(N)^2
\end{rcases}
, p,q \neq 0
\eea
with 
\beq
\hat{\Theta}_{0}(N)=\sqrt{N-\sum_{p\neq 0}\hat{b}^{\dag}_{p}\hat{b}_{p}}
\eeq
Here, $0$ is the reference state and $p,q =\pm1,\pm2$ are the non-zero momentum states of the system. $\hat{b}_{p}$ are the new HP bosonic operators and satisfy the commutation relation - $[\hat{b}_{p},\hat{b}^{\dag}_{q}]=\delta_{p,q}$.
The expectation value for these bosonic modes is $\langle\hat{b}_{p}^{\dag}\hat{b}_{p}\rangle\leq N$ for $p\neq m$. However, all of them are macroscopic in the limit $N \rightarrow \infty$ in order to make the HP approximation valid. 
Substituting these expressions in the Hamiltonian (\ref{2pump1}) , we get -
\bea
\hat{H}_{MB}&=&  -\hbar\bar{\Delta}_{c}\hat{a}^{\dagger}\hat{a}+\sum_{p=\pm1,\pm2}\hbar\omega_{p} \hat{b}^{\dagger}_{p}\hat{b}_{p} \nonumber\\
&+&\frac{\hbar \lambda}{\sqrt{N}}\left(\hat{a}^{\dag}+\hat{a}\right) \sum_{p=\pm1,\pm2}\left(\hat{b}^{\dag}_{p}\hat{\Theta}_{0}(N)+\hat{\Theta}_{0}(N)\hat{b}_{p}\right) \nonumber \\
& & \label{HP}
\eea 

It may be noted that if the number of excited states $p$ is one, in that case we can directly use a pseudospin - $\frac{1}{2}$ representation of these bosonic operators and the last term of 
 the Hamiltonian indicates the coupling of a single bosonic mode with a large single spin $S$. And that is the prototype Dicke Hamiltonian \cite{Lieb, Wang} which has been experimentally verified to show a super-radiant quantum phase transition above a critical cavity-pump detuning \cite{Baumann}. Here, we present a generalisation of the HP approximation because of 
 the existence of more than one excited state. In the subsequent discussion, we shall directly use the Hamiltonian (\ref{2pump1}).
 
It may be pointed out that an optical cavity is characterised by the Purcell factor $\beta$ and the intra-cavity field decay rate $\kappa$. For $\beta > 1$, the scattering into modes not supported by the cavity is practically eliminated. The typical experimental situation that motivates our theoretical proposal ( such as 
\cite{Baumann}) is carried out in the regime $\beta > 1$ and  $\kappa \gg \omega_{\text{rec}}$. This criterion allows the kinetic energy transfer for the backscattering of two photons into different momentum modes  that are 
supported by the cavity. As a result, in this case Holstein-Primakoff approach works. Alternative regime where $\kappa < \omega_{\text{rec}}$ was also explored in experiments \cite{Wolke}.

The Dicke Hamiltonian can be exactly diagonalized in the thermodynamic limit ($N\rightarrow\infty$) using the HP approximation \cite{Emary1,Emary}. 
It shows a continuous transition from a normal to a super-radiant phase in this limit as a function of the critical value of the atom-pump coupling $\lambda$ defined in (\ref{crdeltakappa}). Given the fact that our system contains a finite but large number of particles, application of this HP approximation provides a reasonable value
at which such transition occurs as established by recent experiments \cite{Baumann, Leonard}. The  HP approximation, however, breaks down when there is a superfluid to insulator type of transition and one needs different  method to study such a system \cite{Hemmerich}. We shall discuss this issue in more detail in a 
later Section \ref{BHM}.

We now expand the atomic and cavity field operators using the Holstein-Primakoff transformation \cite{Emary1, Emary}-
\begin{subequations}
\begin{eqnarray}
\hat{a} &=& \sqrt{N}\alpha + \delta\hat{a} \label{HP1} \\
\hat{c}_{1\pm} &=& \sqrt{N}\Psi_{1\pm} + \delta\hat{c}_{1\pm} \label{HP2} \\
\hat{c}_{2\pm} &=& \sqrt{N}\Psi_{2\pm} + \delta\hat{c}_{2\pm} \label{HP3}\\
\hat{c}_{0} &=& \sqrt{N-\hat{c}^{\dag}_{1-}\hat{c}_{1-}-\hat{c}^{\dag}_{1+}\hat{c}_{1+}-\hat{c}^{\dag}_{2-}\hat{c}_{2-}-\hat{c}^{\dag}_{2+}\hat{c}_{2+}} \nn \\ \label{HP4}
\end{eqnarray}
\end{subequations}

\begin{figure*}
\includegraphics[width=1.8\columnwidth, height=0.8\columnwidth]{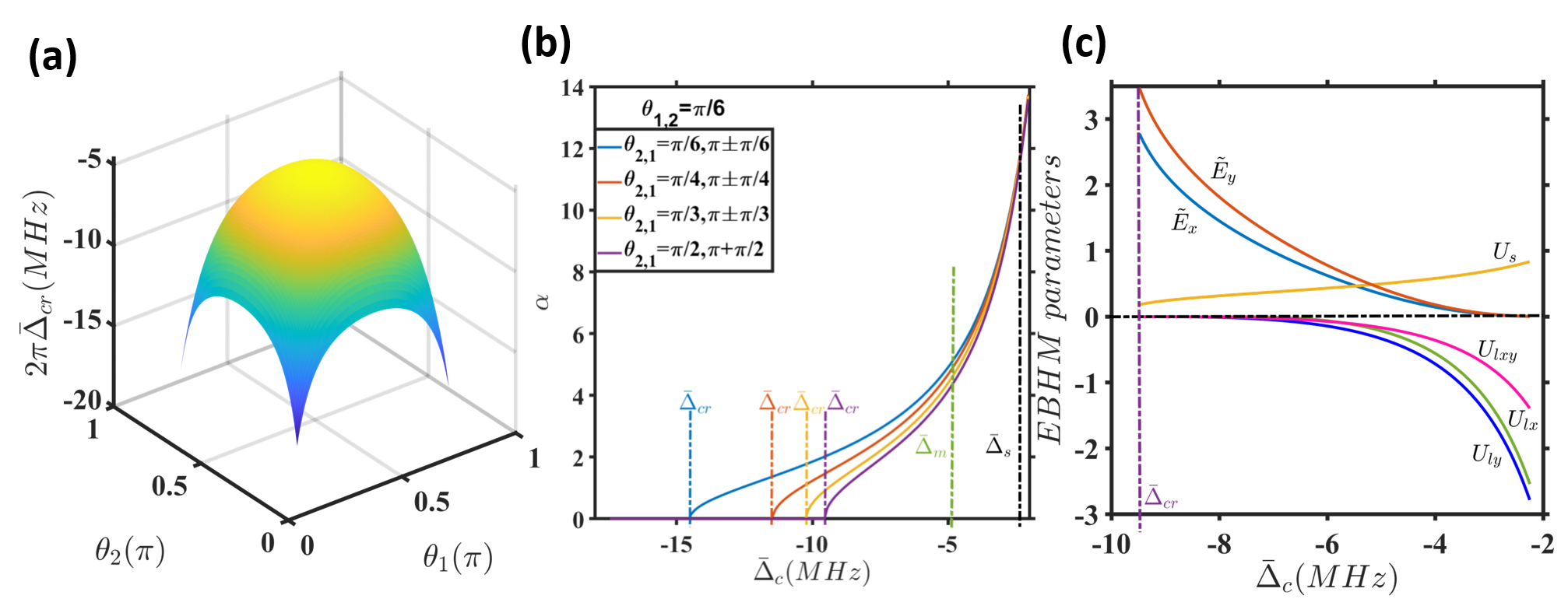}
\caption{(\textit{color online}) (a) The values of the critical detuning as function of $\theta_1$ and $\theta_2$. (b) $\alpha$ as a function of $\bar{\Delta}_c$ for constant $\theta_1$ and different values of $\theta_2$. The notations imply that $\theta_{1}$ and $\theta_{2}$ are interchangeable. $\bar{\Delta}_{cr}, \bar{\Delta}_{m}, \bar{\Delta}_{s}$ are explained in the text. (c) 
EBHM parameters ( for $\theta_{1}, \theta_{2}= \frac{\pi}{6}, \frac{\pi}{2}$) $\tilde{E}_{x,y}$ ( tunneling parameters), the on-site energy $U_{s}$ and the strength of the long-range interaction $U_{lx,ly, lxy}$ in units of $E_{rec}$ in Hamiltonian (\ref{Hbh})
as a function of $\bar{\Delta}_c$. We have rescaled $\tilde{E}_{x,y}$ by a factor of 10 to plot them in the same graph as $U_{s},U_{lx},U_{ly}, U_{lxy}$.} 
\label{alphaEBHM}
\end{figure*}
where the first term in each expansion represents the ground state expectation value and the second term is the fluctuation. 
It may be pointed out that $\psi_{0,1\pm,2\pm}$ are the wave functions for the momentum modes $\vert 0, \bs{k}_c\pm\bs{k}_{1,2}\rangle$ whereas  $\Psi_{0,1\pm,2\pm }$ are the mean field values of $\hat{c}_{0,1\pm,2\pm}$.
Inserting expressions (\ref{HP1}) - (\ref{HP4}) in  Eq.$(\ref{2pump1})$, the many-body Hamiltonian can be split into three parts and is written as 
\beq  
\hat{H}_{MB} = N\hat{h}^{(0)}_{m=0}+\sqrt{N}\hat{h}^{(1)}_{m=0}+\hat{h}^{(2)}_{m=0} \label{HHPG} 
\eeq 
with  each part scaling as $N^{(2-n)/2}$. In the expression (\ref{HHPG}) 
\bea 
\hat{h}^{(0)}_{m=0} &=&-\hbar \bar{\Delta}_c \alpha^2+\hbar\omega_{1+}\Psi^2_{1+}+\hbar\omega_{2+}\Psi^2_{2+} \nn \\ 
&  & \mbox{} +\hbar\omega_{1-}\Psi^2_{1-}+\hbar\omega_{2-}\Psi^2_{2-}  \nn \\
&  & \mbox{} + 4\hbar\lambda\alpha\Psi_{0}(\Psi_{1+}+\Psi_{1-}+\Psi_{2+}+\Psi_{2-}) 
\eea 
and $\hat{h}^{(1)}_{m=0}$ and $\hat{h}^{(2)}_{m=0}$ are respectively linear and quadratic in fluctuations.

\subsection{Ground state properties}\label{gsp}
The ground state energy is obtained from 
$\frac{\partial\hat{h}^{(0)}_{m=0}}{\partial \alpha}=0$, $\frac{\partial\hat{h}^{(0)}_{m=0}}{\partial \Psi_{1\pm}}=0$ and $\frac{\partial\hat{h}^{(0)}_{m=0}}{\partial \Psi_{2\pm}}=0$.

\begin{figure*}
\includegraphics[width=1.8\columnwidth, height=1.5\columnwidth]{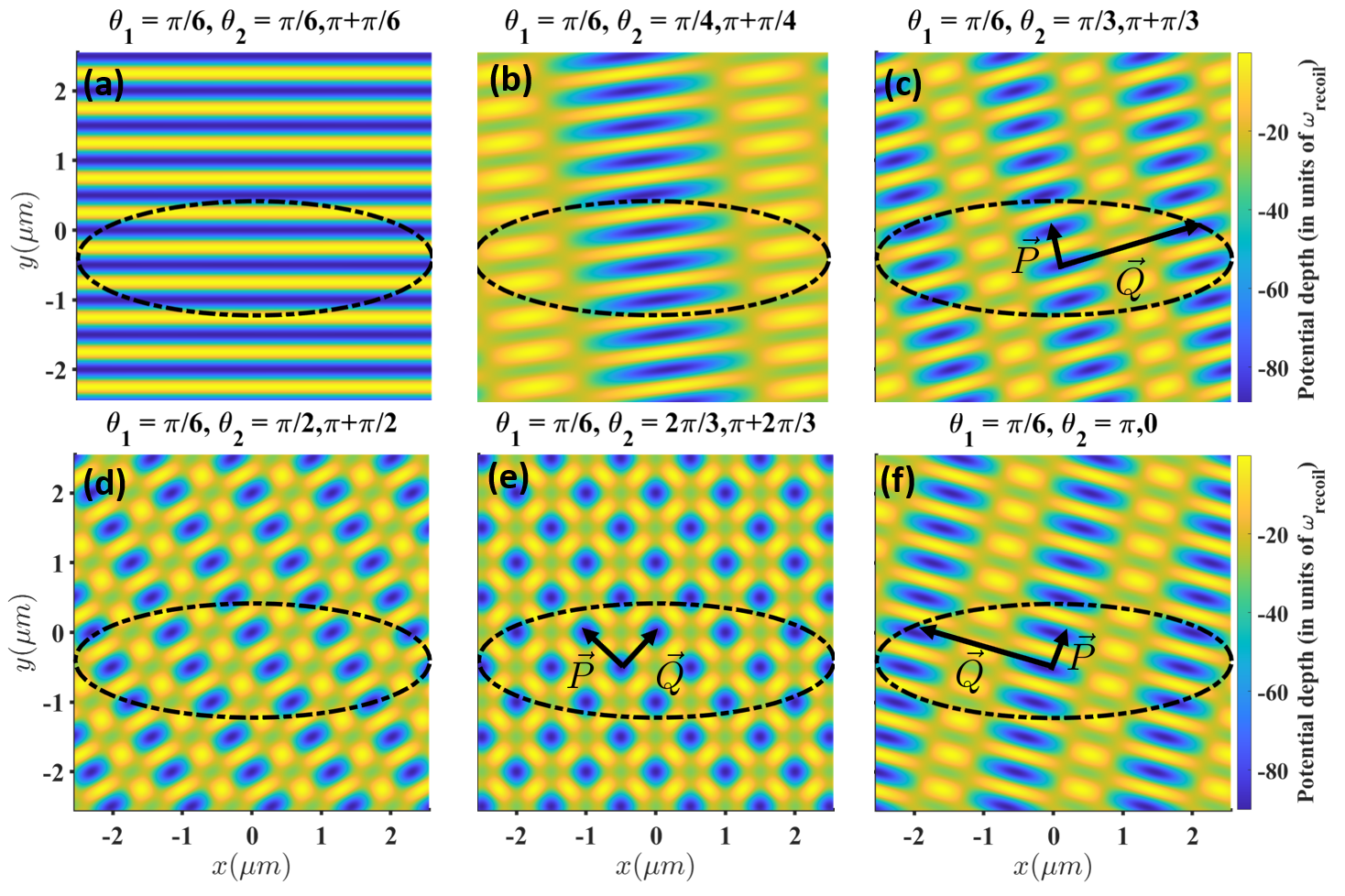}
\caption{(\textit{color online}) (a)-(f) Dynamical optical lattice potential for fixed value of $\theta_1$ and varying $\theta_2$ according to the expression (\ref{Valpha}). Angles are mentioned on the top of each figure. $x$-axis is same in either row. $\vec{P}$ and $\vec{Q}$ are defined in the text. }\label{lattices}
\end{figure*}
\begin{figure*}
\includegraphics[width=1.8\columnwidth, height=1.5\columnwidth]{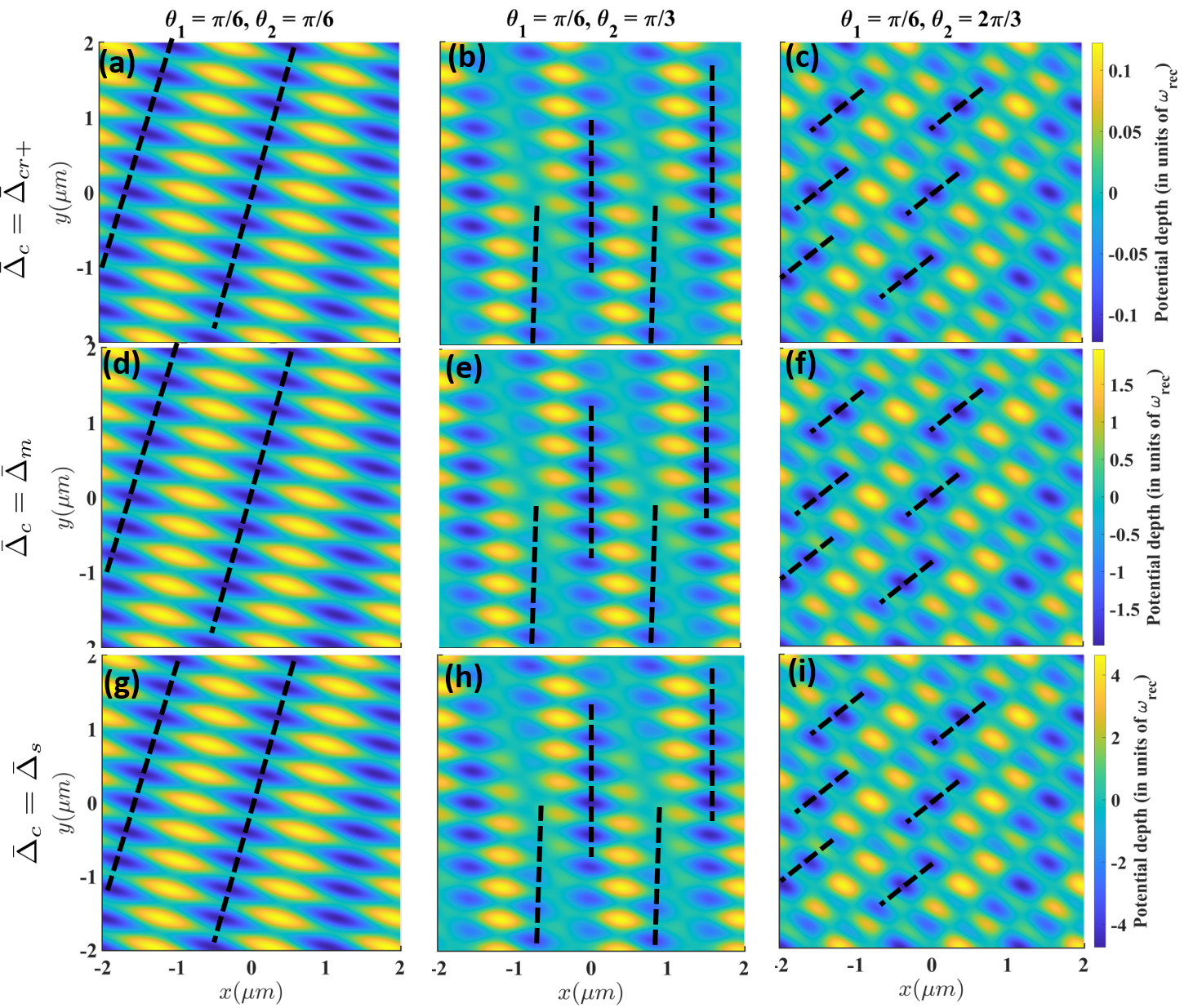}
\caption{(\textit{color online}) (a) -(i) Dynamical optical lattice potential for fixed value of $\theta_1$ and varying $\theta_2$ (indicated on the top of each figure) for different values of $\bar{\Delta}_{c}$ (indicated on the left) according to expression (\ref{Valpha_HP}). 
As $\bar{\Delta}_{c}$ increases, the depth of the potential increases but the structure of the potential remains the same. The dashed black line indicated the locations of potential minima. See details in Section(\ref{sec_QOL}) } \label{OL_HP}
 \end{figure*}
 
For $\alpha$, we obtain -
\beq 
\alpha =\frac{2\lambda}{\bar{\Delta}_c}\Psi_0(\Psi_{1+}+\Psi_{1-} + \Psi_{2+}+\Psi_{2-})
\label{alpha_2pump}
\eeq 
Substituting the value of $\alpha$ in $\hat{h}^{(0)}_{m=0}$ gives -
\bea
\hat{h}^{(0)}_{m=0} &=& \hbar\omega_{1+}\Psi^2_{1+}+\hbar\omega_{1-}\Psi^2_{1-}+\hbar\omega_{2+}\Psi^2_{2+}+\hbar\omega_{2-}\Psi^2_{2-} \nn \\
& + & \frac{4\hbar\lambda^2}{\bar{\Delta}_c}\Psi^2_0(\Psi_{1+}+\Psi_{1-})^2 + \frac{4\hbar\lambda^2}{\bar{\Delta}_c}\Psi^2_0(\Psi_{2+}+\Psi_{2-})^2 \nn \\
&  + & \frac{8\hbar\lambda^2}{\bar{\Delta}_c}\Psi^2_0(\Psi_{1+}+\Psi_{1-})(\Psi_{2+}+\Psi_{2-}) \label{Halpha}
\eea 
The expression (\ref{Halpha}) is now extremized with respect to $\Psi_{1+}$, $\Psi_{1-}$,$\Psi_{2+}$ and $\Psi_{2-}$  which yields four Eqs. (\ref{eqpsi1p}), (\ref{eqpsi2p}),  (\ref{eqpsi1m}),  (\ref{eqpsi2m})
( for details see Appendix \ref{GSEQ}) .
 These are solved simultaneously subject to the conditions 
 \bea
\Psi_{1+}^{2}&+&\Psi_{1-}^{2}+\Psi_{2+}^{2}+\Psi_{2-}^{2}+\Psi^{2}_{0}=1;\\
0&<&\Psi_{1\pm}< 1;\quad 0<\Psi_{2\pm}<1, \eea
to obtain the solutions for $\Psi_{1+}$, $\Psi_{1-}$, $\Psi_{2+}$ and $\Psi_{2-}$. Then these values are substituted in Eq.(\ref{alpha_2pump}) to obtain $\alpha$ as a function of $\bar{\Delta}_c$.
The critical detuning is given as ( for detailed derivation see Appendix \ref{CRD}) -
\beq
\bar{\Delta}_{cr}=-\frac{4\lambda^2}{\bar{\omega}_{1}}-\frac{4\lambda^2}{\bar{\omega}_{2}}\label{crdelta}
\eeq
where $\bar{\omega}_{1}^{-1}=\omega_{1+}^{-1}+\omega_{1-}^{-1}$ and $\bar{\omega}_{2}^{-1}=\omega_{2+}^{-1}+\omega_{2-}^{-1}$. In presence of atom-atom interactions and considering cavity decay rate, $\kappa$, the critical detuning is modified as 
\begin{widetext}
\begin{eqnarray}
\bar{\Delta}_{cr} = -\frac{2\lambda^2 }{\omega_{10}}-\sqrt{\frac{-4\lambda^4 }{\omega_{10}^2}-\kappa^2}-\frac{2\lambda^2 }{\omega_{20}}-\sqrt{\frac{-4\lambda^4 }{\omega_{20}^2}-\kappa^2}\label{crdeltakappa}
\end{eqnarray}
where $\omega_{10}=\left(\frac{g_{2D}\vert\Psi_0\vert^2}{2\hbar}+\bar{\omega}_1\right)$ and $\omega_{20}=\left(\frac{g_{2D}\vert\Psi_0\vert^2}{2\hbar}+\bar{\omega}_2\right)$. 
\end{widetext}

In Fig.\ref{alphaEBHM}(b), we provide a plot of $\alpha$ values as a function of $\bar{\Delta}_c$ for different values of $\theta_1$ and $\theta_2$. As can be seen from Fig.\ref{alphaEBHM}(a) that 
 for $\theta_1=\theta_2$, the critical detuning, $\bar{\Delta}_{cr} $ is maximum and then it decreases symmetrically from the maximum value as $|\theta_{1} - \theta_{2}|$ increases.  
 For fixed value of $\theta_1(\theta_2)$, the critical detuning increases upto $\theta_2(\theta_1)=\pi/2$ and then decreases symmetrically upto $\theta_2(\theta_1)=\pi$. This happens because as $\theta_2$ increases, $\bar{\omega}_2$ increases which results in a decrease in  $\vert\bar{\Delta}_{cr}\vert$ (see Eq.(\ref{crdelta},\ref{crdeltakappa})).  Below $\bar{\Delta}_{cr}$, the system is in the normal phase and $\Psi_{1+}=\Psi_{2+}=\Psi_{1-}=\Psi_{2-}=0$, therefore, $\Psi_{0}=1$ and $\alpha=0$, which represents uniform atomic density. At $\bar{\Delta}_{c}=\bar{\Delta}_{cr}$, the system enters a self-organized supersolid phase, and  $\Psi_{1+}=\Psi_{2+}=\Psi_{1-}=\Psi_{2-}\neq 0$, which results in $\Psi_{0}\neq 1$ and $\alpha\neq 0$. As $\vert\bar{\Delta}_{c}\vert$ decreases, $\alpha$ increases. From Eq.(\ref{eqpsi1p},\ref{eqpsi2p},\ref{eqpsi1m},\ref{eqpsi2m}), we can see that $\Psi_{1\pm}$ and $\Psi_{2\pm}$ depend on $\omega_{1\pm}$ and $\omega_{2\pm}$, respectively. We observe that $\omega_{2+}$($ \pi\pm \theta_2)=\omega_{2-}$($\theta_2)$. This results in $\Psi_{2+}\leftrightarrow \Psi_{2-}$. But this interchange in  $\Psi_{2+}$ and $\Psi_{2-}$ does not affect the value of $\alpha$ in Eq.(\ref{alpha_2pump}) because it depends on the total sum  $\Psi_{1+}+\Psi_{1-}+\Psi_{2+}+\Psi_{2-}$. Therefore, for angles $\pi\pm\theta_2$, we obtain same values of $\alpha$ as pointed out in Fig.\ref{alphaEBHM}(b). 

 Following Eq.(\ref{alpha_2pump}),  at resonance $\alpha$ diverges as $\bar{\Delta}_{c} \rightarrow 0$. Accordingly, all the plots in Fig.\ref{alphaEBHM}(b) for different $\theta_{1,2}$ asymptotically approach the same curves when 
 $\bar{\Delta}_{c}$ approaches $0$. To ascertain the behaviour of the self-organised phases close to this resonance value, we have chosen arbitrarily a value of $\bar{\Delta}_{c}$ close to zero where corresponding values of $\alpha$ for various $\theta_{1,2}$ is practically same within the numerical precision for our computation. We call this as 
 $\bar{\Delta}_{s}$. $\bar{\Delta}_{m}$ is intermediate value of the detuning that lies between $\bar{\Delta}_{cr}$ for various $\theta_{1,2}$ and $\bar{\Delta}_{s}$. In the subsequent section, we shall discuss the self-organised atomic phases for these different values of the detuning $\bar{\Delta}_{c}$. 
 \section{Results and Discussion}\label{RD_sec}
 Using the solutions of Eqs. (\ref{eqpsi1p}), (\ref{eqpsi2p}),  (\ref{eqpsi2p}),  (\ref{eqpsi2p}) along with Eq.(\ref{alpha_2pump})  and 
 Eq.(\ref{atomic_op}), we can now evaluate the dynamical quantum optical lattice and the corresponding atomic density in the self-organised super-radiant phases 
 for different values of $\theta_{1}$ and $\theta_{2}$ and a set of $\bar{\Delta}_{c}$
  \subsection{Quantum Optical Lattice Potential in super-radiant phases:}\label{sec_QOL}
To calculate the dynamical optical lattice potential, we replace $\hat{a}(\hat{a}^{\dag})$ in Eq.(\ref{pot_pump}) by this $\alpha(\alpha^{*})$ and plot $V(\bs{r})$ as a function of $x$ and $y$.
The resulting expression becomes 
\bea
V(\bs{r}) & = & \hbar \eta \cos(\bs{k}_{1}\cdot\bs{r} ) \cos(\bs{k}_{c}\cdot\bs{r})(\alpha+\alpha^*) \nn\\
&+&\hbar \eta \cos(\bs{k}_{2}\cdot\bs{r} ) \cos(\bs{k}_{c}\cdot\bs{r})(\alpha+\alpha^*) \nn\\
&+& \hbar U_{0} \cos^{2}(\bs{k}_{c}\cdot\bs{r})\vert\alpha\vert^2\nn\\
&+& \hbar U_{p}\cos^{2}(\bs{k}_{1}\cdot\bs{r})+\hbar U_{p}\cos^{2}(\bs{k}_{2}\cdot\bs{r}) \nn\\
&+& \hbar U_{p}\cos(\bs{k}_{1}\cdot\bs{r} ) \cos(\bs{k}_{2}\cdot\bs{r})\label{Valpha}
\eea
The optical lattice potentials are thus determined by the mean-field value of the photon creation and annihilation operators, which are determined from the coupled atom-photon dynamics inside the cavity. 
 We define $\bar{\Delta}_{cr+}$ as a value just above $\bar{\Delta}_{cr}$. Under HP approximation $\alpha= \alpha^{*}$, which we determine numerically from Eq.(\ref{alpha_2pump})  and substitute in the expression (\ref{Valpha}) at this detuning $\bar{\Delta}_{cr+}$, and provide a few representative plots for fixed $\theta_1=\pi/6$ and varying $\theta_2$ in Fig.\ref{lattices}. In Fig.\ref{lattices}(a) $\theta_2=\theta_1=\pi/6$, and we obtain stripes of maxima($yellow$) and minima($blue$) forming a one-dimensional potential along the $y-$ direction.
It may be noted that for $\theta_1=\theta_2=\pi/6$, we get $\Psi_{1+}=\Psi_{2+} $ and $\Psi_{1-}=\Psi_{2-}$, which is the solution of an effective single pump-single cavity arrangement  which has an intensity $2U_{p}$ \cite{Leonardthesis}. .
\begin{figure*}
\includegraphics[width=2\columnwidth, height=1.7\columnwidth]{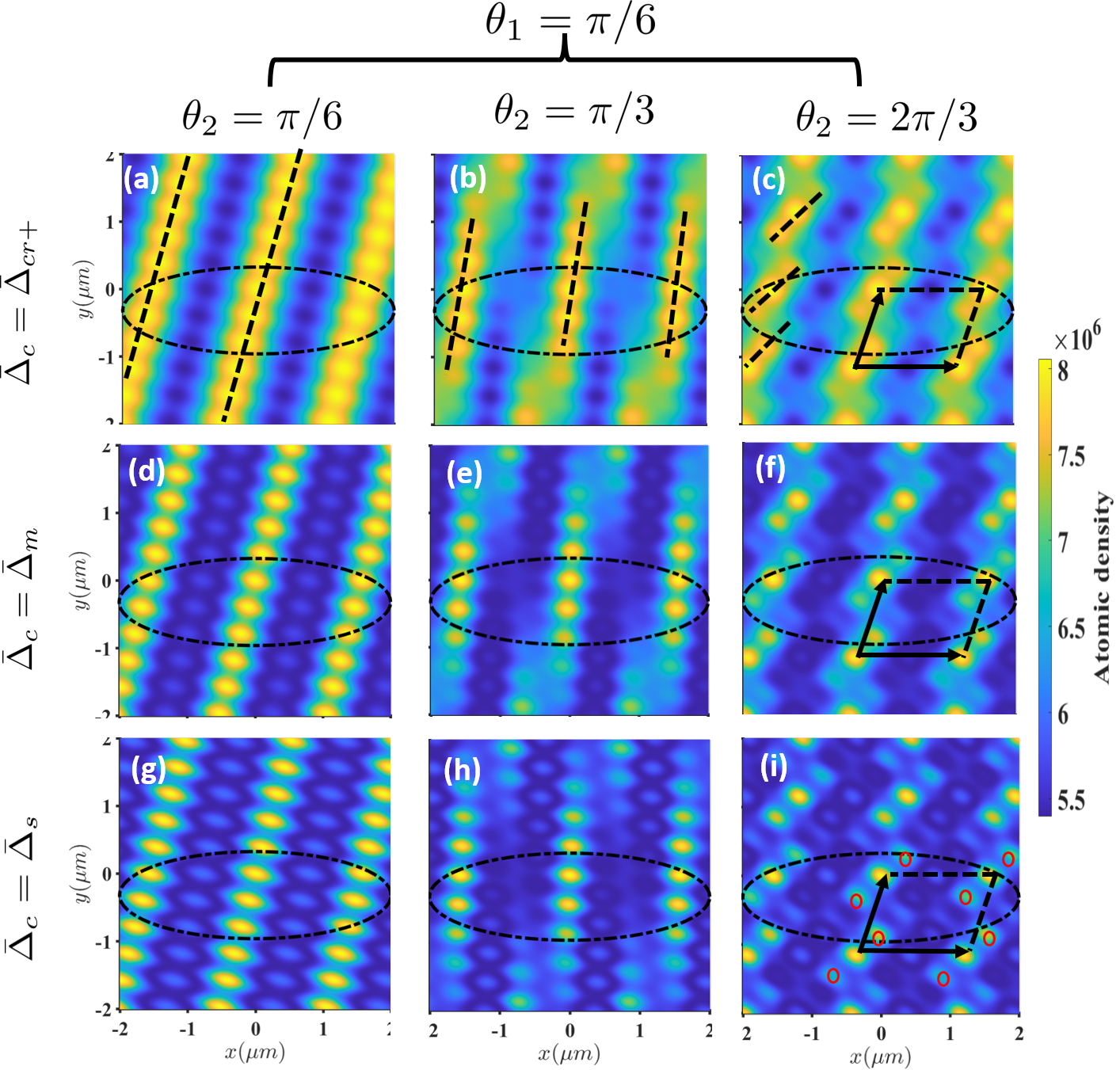}
\caption{(\textit{color online}) (a)-(i)The atomic density (plotted along the color axis) using the Holstein-Primakoff approximation as a function of $x$ and $y$ for $\bar{\Delta}_{cr},\bar{\Delta}_{m}$ and $\bar{\Delta}_{s}$. The sequence of these figures as well as other details are same as the one in Fig.\ref{OL_HP}. The red circles 
in (i) plot indicate the location secondary density maxima in the two-dimensional lattice-supersolid structure and due to the presence of more than one non-zero component of momentum $\bs{k}$ in the superfluid density.
 }\label{pot_atomic1}
\end{figure*}

As we change $\theta_2=\pi/4$ in Fig.\ref{lattices}(b), the minima positions start changing and they gradually start forming a two-dimensional structure. We can understand this by following the top row of the dotted ellipse in these figures. For $\theta_2=\pi/4$, the minima sites of the top row shift towards left and form a rectangular potential with lattice vectors $\vec{P}$ and $\vec{Q}$. They are given as -
$$ \vec{P}=-\lambda_p\frac{\sin(\theta_2-\theta_1)}{1+\cos(\theta_2-\theta_1)}\hat{x}+\frac{\lambda_p}{2}\hat{y}$$\\
$$\vec{Q}=\lambda_p\frac{\sin(\theta_2-\theta_1)}{1+\cos(\theta_2-\theta_1)}\hat{x}+\frac{\lambda_p}{2}\hat{y}$$
As we increase $\theta_2$ from $\frac{\pi}{2}$ to $\frac{2\pi}{3}$ in Fig. \ref{lattices} (e), the minima sites form a rhombic lattice structure. For $\theta_2=\frac{\pi}{3}$ and at $\theta_2=\pi$ in Fig. \ref{lattices} (c) and (f) one gets 
parallelogramic lattices tilted in mutually opposite orientation.  
Exactly same lattice is obtained when
$\theta_2$ is increased in multiples of$\pi$ which shows that there
is a symmetry in the structure of the potential about the pump $1$
($y$)-axis. In case of photon numbers, this symmetry
exists about the axis perpendicular to the cavity
axis. This is because for any change $\pm\Delta\theta$ from this axis, we obtain the
same cavity field amplitude. 

In Fig.(\ref{OL_HP}), we plot  the optical lattice potential that we actually use in the calculation of atomic density in the self-organised phases under HP approximation. The corresponding expression is just the $\alpha$ dependent part in the expression (\ref{Valpha}) and hence gives us the dynamic part of the potential.
 \bea
 V(\bs{r}) & = & \hbar \eta \cos(\bs{k}_{1}\cdot\bs{r} ) \cos(\bs{k}_{c}\cdot\bs{r})(\alpha+\alpha^*) \nn\\
&+&\hbar \eta \cos(\bs{k}_{2}\cdot\bs{r} ) \cos(\bs{k}_{c}\cdot\bs{r})(\alpha+\alpha^*) \nn\\
&+& \hbar U_{0} \cos^{2}(\bs{k}_{c}\cdot\bs{r})\vert\alpha\vert^2 \label{Valpha_HP} \eea 
 The dotted lines in Fig.(\ref{OL_HP})(a)-(i) show the minima sites of the potential and will correspond to the maxima of the atomic density. For each $\theta_{2}$ we plot the optical lattice potentials from top to bottom in increasing order of $\bar{\Delta}_{c}$. As expected the potential is deepest ( see the colour bar) in the lowest row, namely in  Fig.(\ref{OL_HP})(g)-(i). From left to right in each row with increasing $\theta_{2}$,the quantum optical lattice shows a transition from one dimensional form to a two dimensional form. The  corresponding density patterns will be discussed in more details in the next sub-section \ref{sec_dens}.

\subsection{Self-organised atomic density}\label{sec_dens}
As compared to the normal phase the condensate, where the atoms only populate the zero-momentum state in the super-radiant phase, other momentum states depicted in  Fig.(\ref{mom_2pump}) gets populated at different values of the pump  angles $\theta_{1,2}$ leading to a phase transition. To plot the atomic densities in these new phases, we substitute the numerical solutions of Eq.(\ref{eqpsi1p},\ref{eqpsi2p},\ref{eqpsi1m},\ref{eqpsi2m}), in Eq.(\ref{atomic_op}). Then $A\vert \Psi\vert^2$ is plotted as a function of $x$ and $y$, for fixed $\theta_1= \pi/6$ and variable $\theta_2$ in Fig.(\ref{pot_atomic1}) (a)-(i). It may be noted that for  $\bar{\Delta}_{c}<\bar{\Delta}_{cr}$, $\alpha=0$. For $\theta_1= \pi/6=\theta_2$, $\alpha$ becomes non-zero for $\bar{\Delta}_{c}=\bar{\Delta}_{cr}$,  which results in the localization of atoms (Fig.(\ref{pot_atomic1}) (a),(d),(g)) at the minima sites  of the optical lattice potential showing a one-dimensional variation given in Fig.(\ref{OL_HP})(a),(d),(g).
For $\bar{\Delta}_{c}=\bar{\Delta}_{m}$ and $\bar{\Delta}_{c}=\bar{\Delta}_{s}$, this one dimensional localization gets stronger due to increase in $\alpha$.  As we increase $\theta_2$, the arrangement of the minima sites start deviating from this perfect one-dimensionality. Consequently the atoms start relocating them according to the new-pattern of potential minima and form periodic pattern that is intermediate between a one dimensional and two dimensional pattern. This can be 
seen in (Fig.(\ref{pot_atomic1}) (b),(e),(h)). For $\theta_2=\frac{2\pi}{3}$, increasingly two-dimensional arrangement of potential minima are available for occupation ( see Fig.(\ref{OL_HP})(c),(f),(i)) and this gives rise to a prominent two-dimensional variation of the atomic density in Fig.(\ref{pot_atomic1}) (c),(f),(i)) where the unit cell is identified inside each figure. Vertically downward in this column the atomic density increases as $\alpha$ increases with increasing $\bar{\Delta}_{c}$, making the two dimensional structure more prominent. Appearance of such self-organised periodic modulation of the superfluid density above the critical detuning $\bar{\Delta}_{c}$ in a finite system 
is a hallmark of the lattice-supersolid phase \cite{Leonard}. These figures thus show a clear dimensional cross-over in self organized lattice supersolid phases in the superradiant regime and represent the central result of this work. 
Because of the presence of several momentum components in the expression (\ref{atomic_op}), there are also secondary atomic density minima  some of which are marked with red circle (see Fig.\ref{pot_atomic1} (i)).
This self-organisation is an outcome of the cavity-mediated long-range interaction between the atoms. The explicit form of this long-range interaction appears clearly in an Extended Bose-Hubbard model (EBHM) derived under self-consistent tight-binding approximation in this quantum optical lattice potential. In the next section, we will derive this  EBHM for this system and relate the BH parameters with the obtained $\alpha$ values and the dynamical optical lattice potential obtained under Holstein-Primakoff transformation. 

\section{Classification of the super-radiant phases}\label{class}
The solutions of Eqs.(\ref{eqpsi1p}), (\ref{eqpsi2p}),  (\ref{eqpsi2p}),  (\ref{eqpsi2p})  yield for $\bar{\Delta}_{c} < \bar{\Delta}_{cr}$, 
$\Psi_{1\pm}=\Psi_{2\pm}=0$, which gives, $\Psi_{0}=1$ and the atomic density $\vert \Psi(x,y) \vert^2=1/A$.  This characterises a homogenous superfluid phase (HSF) in the normal region. 
In this phase, the cavity photon number is zero as shown in Fig.\ref{alphaEBHM}(b). For  $\bar{\Delta}_{c}>\bar{\Delta}_{cr}$,  
$$\Psi_{1+}=\Psi_{2+}=\Psi_{1-}=\Psi_{2-}\neq 0,$$ which gives, $\Psi_{0}\neq 1$ and the system enters a super-radiant phase with the appearance of output cavity photons making $\alpha \neq 0$.

\begin{figure*}
\includegraphics[width=2\columnwidth, height=2\columnwidth]{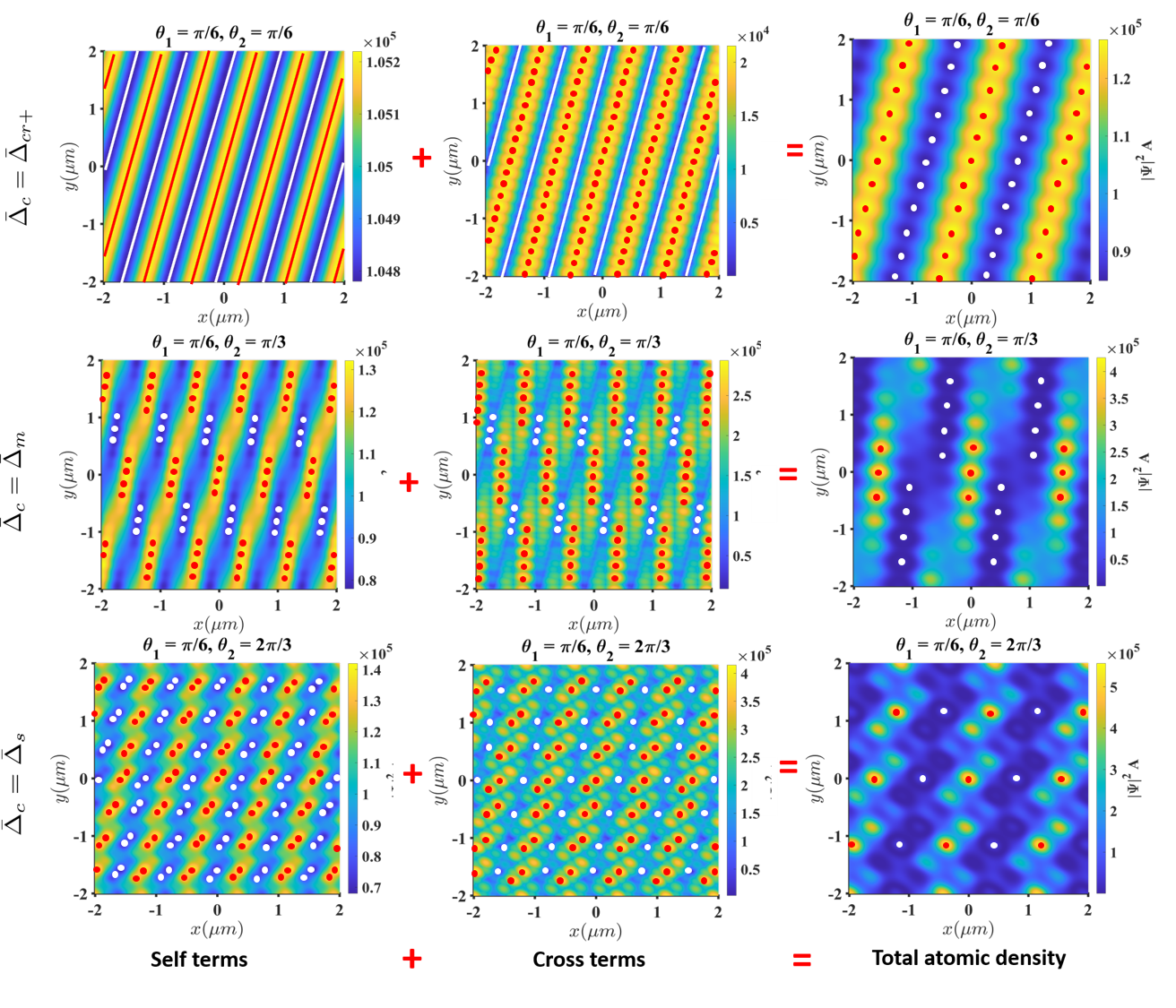}
\caption{(\textit{color online}) We plot the self -interference terms and cross-interference terms of the atomic density for different cavity-pump detunings, $\bar{\Delta}_{cr+}, \bar{\Delta}_{m}$ and $\bar{\Delta}_{s}$ as given in the expression (\ref{atomicden}). The atomic density (plotted along the color axis) is calculated using the Holstein-Primakoff approximation as a function of $x$ and $y$. The $red$ dots show the maxima points and the $white$ dots show the minima points of the atomic density. The pump angles corresponding to each figure is shown on the top of each figure.}  \label{S_C}
\end{figure*}

The atomic density can be obtained by substituting the mean field part of each operator from Eqs. (\ref{HP1}) -(\ref{HP4}) in the expression of the atomic operator 
(\ref{atomic_op}), namely 
\begin{widetext}
\bea
\vert\Psi\vert^2 & = & \vert\psi_0\Psi_0 + \psi_{1+}\Psi_{1+}+\psi_{1-}\Psi_{1-}+\psi_{2+}\Psi_{2+}+\psi_{2-}\Psi_{2-}\vert^2 \nn \\
&=& \vert\psi_{0}\Psi_{0}\vert^2 + \vert\psi_{1-}(x,y)\Psi_{1-}\vert^2 + \vert\psi_{1+}(x,y)\Psi_{1+}\vert^2+ \vert\psi_{2-}(x,y)\Psi_{2-}\vert^2 + \vert\psi_{2+}(x,y)\Psi_{2+}\vert^2 \nn\\
&+& 2\psi^*_{0}\psi_{1-}\Psi^{*}_{0}\Psi_{1-}+
2\psi^*_{1-}\psi_{1+}\Psi^{*}_{1-}\Psi_{1+}+
2\psi^*_{1+}\psi_{2-}\Psi^{*}_{1+}\Psi_{2-}+
2\psi^*_{2+}\psi_{2-}\Psi^{*}_{2+}\Psi_{2-}\nn\\&+&
2\psi^*_{0}\psi_{2+}\Psi^{*}_{0}\Psi_{2+}+
2\psi^*_{0}\psi_{1+}\Psi^{*}_{0}\Psi_{1+}+
2\psi^*_{0}\psi_{2-}\Psi^{*}_{0}\Psi_{2-}+
2\psi^*_{2-}\psi_{1-}\Psi^{*}_{2-}\Psi_{1-}\label{atomicden}
\eea
\end{widetext}
The first five terms of the resulting expression are proportional to the single-mode density $|\Psi_{i \pm}|^{2}$ with $i=0,1,2$. We refer them as the self terms. 
The other terms contain the overlap of such single mode superfluid order parameter at a specific $\bs{k}$, and are proportional to $\Psi_{i\pm} \Psi_{j\pm}$ with $i \neq j$ and $i,j=0,1, 2$. We refer such terms as cross terms.
Our calculation shows that the either type of terms significantly contribute towards the formation of a self-organised lattice structure in the super-radiant regime.
We provide explicitly the contribution of the self atomic density in Fig.\ref{S_C} for $\bar{\Delta}_{cr+}, \bar{\Delta}_{m}$ and $\bar{\Delta}_{s}$, 
for the  same representative combination of pump angles $\theta_{1,2}$ as the ones considered in Fig.\ref{pot_atomic1}. 

We use the maxima and minima points of these atomic density plot to identify the resulting self-organised lattice  structures in the super-radiant regime. To that purpose we first set $F(x,y)=\vert\Psi(x,y)\vert^2$
The extrema points of the atomic density can be obtained from the conditions 
\bea F_{x} = \frac{\partial F}{\partial x} &= & 0 \nn\\
 F_{y} = \frac{\partial F}{\partial y} &=&  0 \nn \eea. 
 This gives us two equations in two unknown variables, $x$ and $y$, and their solution will give us the extrema points of the atomic density. Maxima and minima of $F(x,y)$ can be determined from
\beq F_{xx}=\frac{\partial^2 \vert\Psi(x,y)\vert^2}{\partial x^2}, F_{yy}=\frac{\partial^2 \vert\Psi(x,y)\vert^2}{\partial y^2}, F_{xy} = \frac{\partial^2 \vert\Psi(x,y)\vert^2}{\partial x\partial y}. \nn \eeq 
For $ F_{xx} F_{yy}-F_{xy}>0$, the solution $F(x,y)$ can be a maximum or a minimum point. Then, if $F_{xx}<0$ and $F_{yy}<0$, then $(x,y)$ is a maximum point and if $F_{xx}>0$ and $F_{yy}>0$, then $(x,y)$ is a minimum point. If $ F_{xx}F_{yy}-F_{xy}<0$, then $(x,y)$ is a saddle point. In general such points have to be obtained through numerical computation. But for some specific combination of angles we can determine the maxima and minima points from an analytically solvable equations. For example for $\theta_{1}=\theta_{2}$, where the  two-pump condition degenerates into a single pump case 
The $x$ and $y$ co-ordinates for maxima and minima points are -
\bea
x_{\text{max}} &=& \frac{\lambda_p((n+m)-(n-m)\cos\theta_1)}{2\sin\theta_1}\nn\\
y_{\text{max}} &=& \frac{(n-m)\lambda_p}{2}\nn \\
x_{\text{min}} &=& \frac{\lambda_p((n+m+1)-(n-m)\cos\theta_1)}{2\sin\theta_1}\nn\\
y_{\text{min}} &=& \frac{(n-m)\lambda_p}{2}\nn \eea
where $n=0,\pm1,\pm2,\pm3,...$ 
Details of these calculations are provided in the Appendix \ref{MAXMIN}.    

\subsection{Momentum diagram from the atomic density}\label{sec_TOF}
Self-organisation in super-radiant regime is manifested by the sudden-build up of the cavity field accompanied by the formation of the momentum peaks in the absorption image of the atomic cloud after its sudden release \cite{Baumann}. An idea about these momentum peaks  can be obtained by 
taking the Fourier transform (FT) of the atomic density that was analysed in section \ref{class}.   We take the Fourier transform of Eq.(\ref{atomicden}), 
\beq 
\mathcal{F}(k_x,k_y) = \int^{\infty}_{-\infty} \int^{\infty}_{-\infty} e^{2\pi i \bs{k}\cdot \bs{r}}\vert \Psi(x,y)\vert^2 d \bs{r}
\eeq
For the self terms in the expansion (\ref{atomicden}) it gives 
\beq
\int^{\infty}_{-\infty} \int^{\infty}_{-\infty} e^{2\pi \bs{k} \cdot \bs{r}}\vert \psi_{j\pm}\Psi_{j\pm}\vert^2 d \bs{r} =\delta(k_x-K_{xj\pm})\delta(k_y-K_{yj\pm})
\eeq
where $j=0,\pm{1}, \pm{2}$, $K_{xj\pm}$ is the $x$-component of $\vec{k}_{c}\pm\vec{k}_{j}$ and $K_{yj\pm}$ is the $y$-component of $\vec{k}_{c}\pm\vec{k}_{j}$.
Similar expressions can also be obtained by the cross-terms. 
%
%

\begin{figure*}
\includegraphics[width=2\columnwidth, height=1.6\columnwidth]{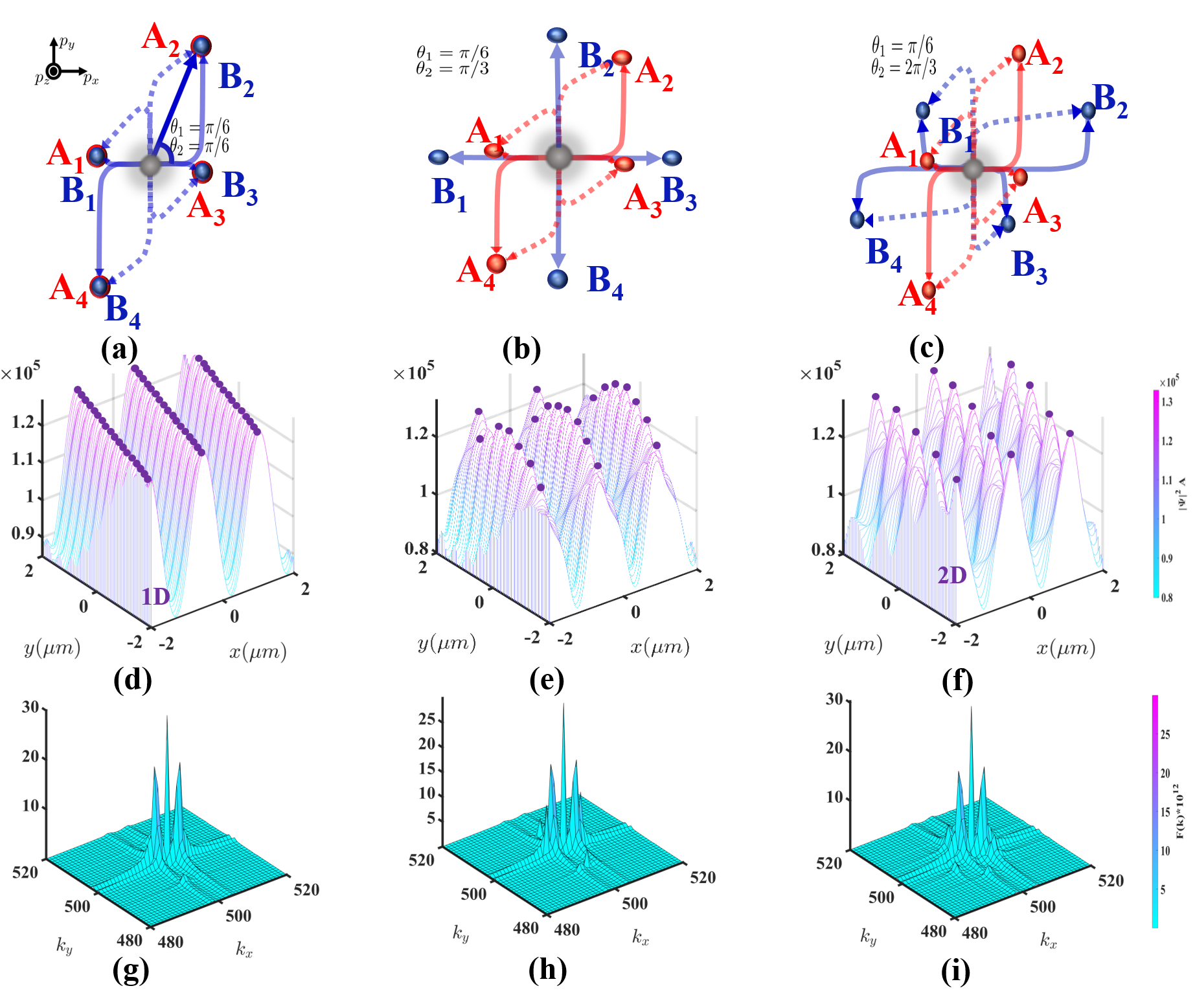}
\caption{(\textit{color online}) (a)-(c) The momentum scattering diagrams for three $(\theta_1,\theta_2)$ combinations. (d)-(f) The waterfall plot of the atomic density is shown which clearly shows the transition of a 1D lattice supersolid at $\theta_1=\pi/6,\theta_2=\pi/6$ to a 2D lattice supersolid at $\theta_1=\pi/6,\theta_2=2\pi/3$. The maxima density points (shown by $\textbf{purple}$ dots) are well separated along the $x$-direction. The separation between the successive secondary maxima along the $y$-direction is negligible as compared to the separation of the local minima along the $x$-direction. This forms 1D tubes of atoms extending along the $y$-direction. As we change the value of pump-angle $\theta_2$ to $\pi/3$, we can see a separation of the maxima points along the $y$-direction as well. The peak value of the maxima atomic density also shows a variation as compared to the 1D case where all the maxima points have the same peak value. This indicates the onset of the transition of the 1D lattice supersolid to a 2D supersolid. For $\theta_2=2\pi/3$, the maxima atomic density points have equal heights and a comparable separation along the $x$- and the $y$-directions. Therefore, this arrangement can be identified as a 2D lattice supersolid. (g)-(i) We plot the Fourier transform of the atomic density for $\bar{\Delta}_{cr+}$.  The additional peaks in these plots are due to the presence of secondary density maxima points (shown in Fig.(\ref{pot_atomic1})) which are a result of the cross-interference terms shown in Fig.(\ref{S_C}). } \label{FFT_waterfall}
 \end{figure*}

In Fig.(\ref{FFT_waterfall}), we show these FT of the atomic densities in the $k_x$ and $k_y$ plane for representative values of $\theta_{1}$ and $\theta_{2}$ along with the 
corresponding momentum scattering diagram and the real space lattice structure of the super-radiant phases. The top row, namely Fig.\ref{FFT_waterfall}(a), (b), (c) plots the momentum-scattering diagram, whereas the bottom row Fig.\ref{FFT_waterfall}(g), (h), (i) shows 
the FT of the atomic density and their peaks. The middle row, thatis Fig.\ref{FFT_waterfall}(d), (e), (f) depicts the maxima point in the real space density by which one can identify the self-organised lattice structure.
The central peak in the momentum distribution at $k_{x0}=k_{y0}=0$, is scaled by a factor of $30$ to show it alongside the rest of the momentum peaks which appear due to the interference of other momenta values and are not captured in Fig.\ref{FFT_waterfall}(a),(b),(c). 

\begin{table*}
\begin{tabular}{ |p{4.0cm}||p{3.0cm}|p{3.0cm}|p{3.0cm}|p{3.0cm}|  }
\hline
 \textbf{$\theta_1=\pi/6$} & \textbf{$(k_x,k_y)_{1+}$} &\textbf{$(k_x,k_y)_{1-}$} &\textbf{$(k_x,k_y)_{2+}$} &\textbf{$(k_x,k_y)_{2-}$}\\
 \hline
 $\theta_2=\pi/6$ & $(k/2 , (1+\sqrt{3}/2)k)$ & $(k/2, (-1+\sqrt{3}/2)k)$ & $(k/2,(1+\sqrt{3}/2)k)$ & $(k/2,(-1+\sqrt{3}/2)k)$\\ 
 $\theta_2=\pi/3$ & $(k/2, (1+\sqrt{3}/2)k)$& $(k/2, (-1+\sqrt{3}/2)k)$ & $(0k,\sqrt{3}k)$ & $(1k,0k)$ \\
 $\theta_2=2\pi/3$ & $(k/2, (1+\sqrt{3}/2)k)$ & $(k/2, (-1+\sqrt{3}/2)k)$ & $(-k/2,\sqrt{3}k/2)$ & $(3k/2,\sqrt{3}k/2)$ \\
 \hline
\end{tabular}
\captionof{table}{\small The position of the prominent peaks ( other than the central peak at $k_{x}=k_{y}=0$), which appear due to the self-terms in the momentum diagram for three combinations of angle $\theta_{1}, \theta_{2}$ as given in Fig. \ref{S_C} ( a subset of the cases presented in Fig. \ref{pot_atomic1}) .}
\label{table_self}
\end{table*}
The  peak locations in the $k_{x}, k_{y}$ plane corresponding to the self-terms for  all three combinations of $\theta_{1}, \theta_{2}$, are presented in a tabular form in Table \ref{table_self}.
These momentum peaks are linked with lattice spacing in the 1D and 2D lattices obtained in Fig.\ref{pot_atomic1} and Fig.\ref{S_C} by the formula $k=2\pi/\lambda_p$. To show that,  we find out the lattice spacing of the 1D lattice in Fig.\ref{pot_atomic1}(a). The lattice spacing for this case is - $\frac{\lambda_p}
{2\cos\frac{\theta_1}{2}}$ which corresponds to a momentum wave vector $\approx 2k$. It can now be checked that the corresponding spacing between the momentum peaks in Table 1 for $(k_x,k_y)_{1+}$ and $(k_x,k_y)_{1-}$ is  $2k$. Similarly, the spacing between the peaks at $(k_x,k_y)_{2+}$ and $(k_x,k_y)_{2-}$ is also $2k$. This demonstrates the relation between the real space lattice and the momentum space diagram.  However in general the presence of multiple momentum peaks lead to a more complex lattice structure whose shape has to be obtained numerically. 

As can be seen in Fig.(\ref{FFT_waterfall})(d) for $\theta_1=\pi/6,\theta_2=\pi/6$, the density maxima points all having same height are very closely spaced along $y$-axis whereas they are well-separated along the $x$-axis. We call such parallel tube like high density regions as one-dimensional lattice supersolid.
In Fig.(\ref{FFT_waterfall})(e) 
for $\theta_1=\pi/6,\theta_2=\pi/3$, some of the maxima peaks shifted from the one-dimensional structure and shows the intermediate stage of a one-dimensional to two-dimensional transition in the structure of the lattice supersolids.
In Fig.(\ref{FFT_waterfall})(f)  for $\theta_1=\pi/6,\theta_2=2\pi/3$, the atomic density maxima are separated almost equally along $x$ and $y$ axis giving a two-dimensional supersolid. As can be seen from the lower panel that the corresponding momentum peaks also change. 

\section{Bose-Hubbard Model for two-pump system}\label{BHM}
Since a dynamical optical lattice is formed inside the cavity, following standard procedure we expand the atomic field operator using the site-localized Wannier functions \cite{Zoller,Fisher, Sheshadri} as
\beq
\hat{\Psi}(x,y) = \sum_{\substack{p,q}}\hat{b}_{p,q} w_{p,q}(x,y) \label{eq3}
\eeq
where $\hat{b}_{p,q}(\hat{b}^{\dag}_{p,q})$ annihilates(creates) an atom at site $(p,q)$ of the cavity,  and $w_{p,q}(x, y)$ is the corresponding maximally localized wave-function. 
It may be noted that these Wannier functions themselves are dynamic since they depend on $\alpha$ \cite{Ritsch, Larson2, Morigi2, Hofstetter}. 

Using the properties of these Wannier function following standard procedure an effective Bose-Hubbard Hamiltonian for the system can be derived as 
\begin{widetext}
\bea
\hat{H}_{BH} &=& \left[E_{x}\hat{B}_{x} + E_{y}\hat{B}_{y} -\hbar(\Delta_{c}-U_0 J_0\hat{N}-U_0\hat{\delta})\hat{a}^{\dagger}\hat{a} 
 +\hbar\eta(\hat{a}^{\dagger} + \hat{a})\left((\tilde{J}_{x1}+\tilde{J}_{x2})\hat{B}_{x}+(\tilde{J}_{y1}+\tilde{J}_{y2})\hat{B}_{y}+(\tilde{J}_{01}+\tilde{J}_{02})\hat{N}\right) \right.\nonumber\\
&&\left.+ \frac{U_{s}}{2}\sum_{p,q}\hat{n}_{p,q}(\hat{n}_{p,q} -1) - \bar{\mu}_{0}\hat{N}\right]  \label{HBH2pump}
\eea
\end{widetext}
It may be noted that the hopping amplitudes along the $x$ and $y$ direction $E_{x},E_{y}$ and the onsite energies $E_{0}$  whose expressions are given in Appendix \ref{expression}, are directly due to the transverse pumping 
The other set of hopping and on-site interactions $(J_{x},J_{y},J_{0},\tilde{J}_{01},\tilde{J}_{02},\tilde{J}_{x1},\tilde{J}_{x2},\tilde{J}_{y1},\tilde{J}_{y2})$ are due to the photons scattered by the atoms and are respectively given in Appendix \ref{expression} and $U_{s} = g_{2D}\int \int dxdy \vert w_{p,q}(x,y)\vert^4 $ is the on-site interaction strength between the atoms. 
$\hat{B}_{x} = \sum_{p,q}(\hat{b}_{p,q}^{\dagger}\hat{b}_{p+1,q} + \hat{b}_{p+1,q}^{\dagger}\hat{b}_{p,q})$ $\hat{B}_{y} = \sum_{p,q}(\hat{b}_{p,q}^{\dagger}\hat{b}_{p,q+1} + \hat{b}_{p,q+1}^{\dagger}\hat{b}_{p,q})$ represents long-range hopping along $x$, $y$ direction. Here $(p+1,q)$ refers to a site along the x-direction and $(p,q+1)$ refers to a site along the y-direction. 
\bea 
\hat{\delta} & = & J_{x}\hat{B}_{x} + J_{y}\hat{B}_{y}\nonumber \\
\hat{n}_{p,q} & = & \hat{b}_{p,q}^{\dagger}\hat{b}_{p,q}, ~~~\hat{N}  =  \sum_{p,q}\hat{n}_{p,q} \nonumber \\
 \bar{\mu}_{0} & = & \mu_0-E_{0} \nn
 \eea

For the parameters considered in this work, $\kappa>>\omega_{rec}$, which refers to the bad cavity limit. In this limit, the cavity decay rate, $\kappa$ is the fastest time scale. $\kappa = 1/\tau$, where $\tau$ is the photon storage time or the total time the photon spends in the cavity. For large values of $\kappa$, $\tau$ is small and as a result, the photons don’t stay inside the cavity for long times \cite{Bakhtiari}. Therefore, the cavity field reaches a steady state well before the atoms. This allows us to adiabatically eliminate the cavity field dynamics by setting 
$i\hbar\frac{\partial \hat{a}}{\partial t}=0$ and obtain the extended Bose-Hubbard Hamiltonian in terms of the atomic operators only. The corresponding expression of $\hat{a}$ is \cite{Maschler} -
\begin{widetext}
\begin{eqnarray}
\hat{a}=\frac{\eta(\tilde{J}_{x1}+\tilde{J}_{x2})\hat{B}_{x}+\eta(\tilde{J}_{y1}+\tilde{J}_{y2})\hat{B}_{y}+\eta(\tilde{J}_{01}+\tilde{J}_{02})\hat{N}}{\Delta_c - U_0J_0\hat{N}-U_0\hat{\delta}+i\kappa}
\end{eqnarray}
For a fixed number of atoms, $N=\langle \hat{N}\rangle$, we expand $\hat{a}$ in tunnelling matrix elements, $J_x$ and $J_y$ as follows 
\begin{eqnarray}
\hat{a}&=&\frac{\eta(\tilde{J}_{x1}+\eta\tilde{J}_{x2})\hat{B}_{x}+\eta(\tilde{J}_{y1}+\tilde{J}_{y2})\hat{B}_{y}+\eta(\tilde{J}_{01}+\tilde{J}_{02})N}{\bar{\Delta}_c-U_0\hat{\delta}+i\kappa}\nn\\
&=& \frac{\eta(\tilde{J}_{x1}+\tilde{J}_{x2})\hat{B}_{x}+\eta(\tilde{J}_{y1}+\tilde{J}_{y2})\hat{B}_{y}+\eta(\tilde{J}_{01}+\tilde{J}_{02})N}{\left(\bar{\Delta}_c+i\kappa\right)}\left(1+\frac{U_0(J_x\hat{B}_x+J_y\hat{B}_y)}{\bar{\Delta}_c+i\kappa}+...\right)\label{alpha_ss}
\end{eqnarray}
and retain only upto the first order terms in the expansion. 
\end{widetext} 
This truncated steady state solution of Eq.(\ref{alpha_ss}) is substituted in Eq.(\ref{HBH2pump}) and we retain terms of second order in $\hat{B}_x$ and $\hat{B}_{y}$ to get the effective EBHM Hamiltonian
as 
\begin{widetext}
\begin{eqnarray}
\hat{H}_{BH}&=& \frac{U_{s}}{2}\sum_{p,q}\hat{n}_{p,q}(\hat{n}_{p,q} -1)- \bar{\mu}_{0}\hat{N}+\tilde{E}_x\hat{B}_x  +\tilde{E}_y\hat{B}_y+U_{lx}\hat{B}_{x}^{2}+U_{ly}\hat{B}_{y}^{2} +U_{lxy}\hat{B}_{x}\hat{B}_{y}+\mathcal{O}(3)\label{Hbh}
\end{eqnarray}
\end{widetext}
We provide a comparison of the above obtained EBHM with other EBHM models studied in \cite{Maschler, Landig, Dogra, Ritsch} in Appendix \ref{Rel_EBHM}.
The detailed expressions of the parameters that appear in the EBHM Hamiltonian in (\ref{Hbh}) are given in Appendix \ref{expression}. 
It is possible to determine the quantum phases (ground state) associated with such effective EBHM Hamiltonian using sophisticated numerical techniques such as dynamical mean-field theory and for a single pump system such work has been done to obtain strongly correlated lattice super-solid phases 
in such systems \cite{Hofstetter}.  Extending such exercise for the current EBHM Hamiltonian (\ref{Hbh}) in our two-pump model  is computationally demanding and is out of scope for the current work. Hopefully this can be explored in future investigations. Before discussing the EBHM further, in the following 
paragraphs, we shall provide a brief discussion about when and why such 
Bose-Hubbard approximation is useful to describe such cavity based ultra cold system as compared to HP approximation described in Section \ref{HPA}.
\begin{figure*}
\includegraphics[width=1.6\columnwidth, height=0.8\columnwidth]{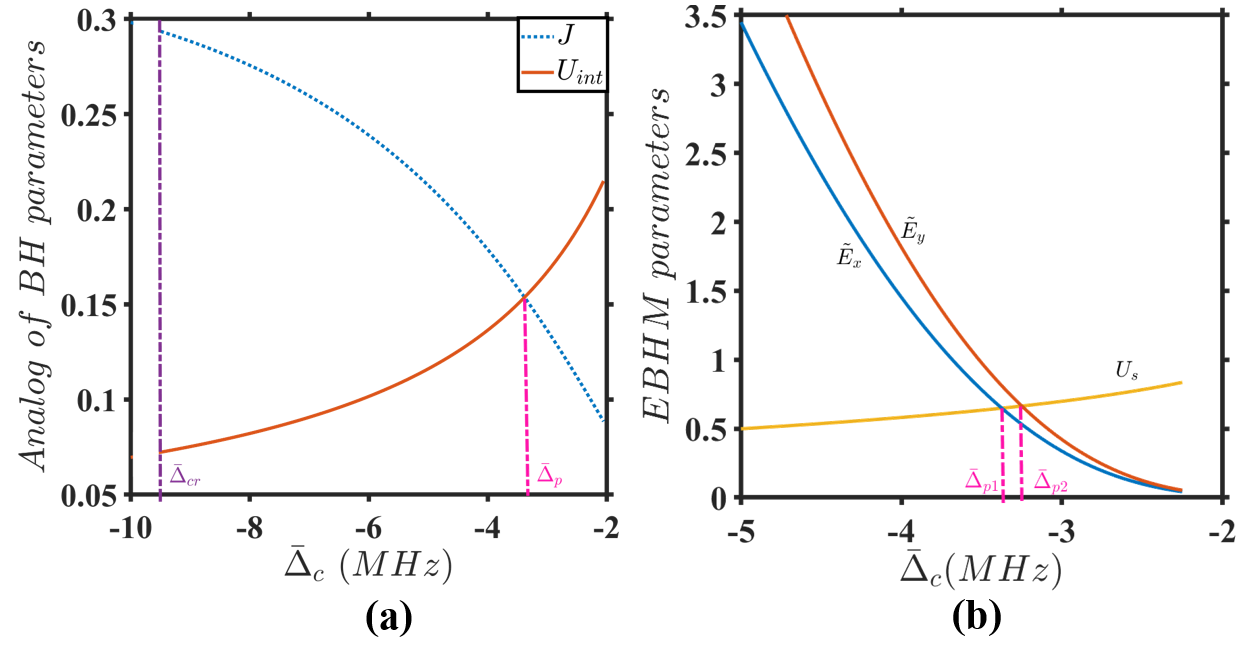}
 \caption{(\textit{color online}) \small (a) Plot of tunnelling amplitude $J$ and on-site interaction strength $U_{int}$ (in the unit of $E_{rec}$) evaluated with the help expression (\ref{classicalBH}) as a function of the detuning parameter. (b) $\tilde{E}_{x,y}$ and $U_s$  (in units of $E_{rec}$), namely the parameters for the EBHM hamiltonian defined in (\ref{Hbh}) plotted as a function of the detuning parameter. The crossings are marked in both plots. 
 It may be noted that in Fig.(\ref{alphaEBHM})(c), the parameters $\tilde{E}_{x,y}$ were scaled by a factor of 10 to show them in the same plot along with the other EBHM parameters. Here in plot(b) we have plotted these parameters as it is. 
}\label{comp_ExEy}
\end{figure*}
It is well known that Bose-Hubbard (BH) Hamiltonian was successfully used to describe the superfluid to Mott insulator transition in ultra cold atomic system \cite{Zoller} in classical optical lattice. In such systems, 
the phase fluctuations, $(\Delta \phi)$, and the number fluctuations, $\Delta N$, follow the uncertainty relation \cite{Pethick, Stringari} - 
\begin{eqnarray}
\Delta N\Delta \phi=1\label{uncert}
\end{eqnarray}
The tunnelling amplitude $J$ and the 
on-site interaction strength $U_{int}$ can be given by the following well known analytical formula valid for a classical optical lattice potential with depth $V_0$ much greater than the recoil energy $E_{rec}$ ($V_0>>E_{rec}$) \cite{Zwerger1} -
\begin{eqnarray}
J&=& \frac{4}{\sqrt{\pi}}E_{rec}\left(\frac{V_0}{E_{rec}}\right)^{3/4}\exp\left(-2\sqrt{\frac{V_0}{E_{rec}}}\right)\nn\\
U_{int}&=&\sqrt{\frac{8}{\pi}}ka_sE_{rec}\left(\frac{V_0}{E_{rec}}\right)^{3/4} \label{classicalBH}. 
\end{eqnarray}  
With the increase in $V_{0}$, the tunnelling of atoms between the minima sites of the optical lattice decreases. This in turn decreases the particle number fluctuations which result in an increase in the phase fluctuations as seen in the uncertainty relation (\ref{uncert}). This leads to the loss of phase coherence and the eventual emergence of a phase- incoherent Mott insulator phase in place of a phase-coherent superfluid phase. 

In the system under consideration, instead of a classical optical lattice, a dynamic quantum optical lattice is formed inside a cavity. Nevertheless, in  Fig.(\ref{comp_ExEy})(a), we  have used the expressions defined in (\ref{classicalBH}) to evaluate the analog quantities of BH parameters for such dynamic 
quantum optical lattice potential. To that purpose, we consider the  $V_0$ to be the depth of the optical lattice potential in Eq.(\ref{Valpha}), and have used later expressions $\alpha(\alpha^*)$ obtained from (\ref{QPBH}). For the system under consideration in this work $E_{rec}$ is defined in Section \ref{HPA} below Eq. (\ref{2pump1}). 
Using this value, we see a crossing of $J$ and $U_{int}$ at $\bar{\Delta}_{p}=-3.35$ $MHz$. Beyond this cavity-pump detuning, the analogue of on-site interactions  in prototype BH model, calculated with the help of quantities defined for the current cavity-atom system, 
 start dominating the corresponding tunnelling between adjacent wells.
 
These results, plotted in Fig.(\ref{comp_ExEy})(a), are now compared with the similar quantities that appear in the EBHM Hamiltonian (\ref{Hbh})  derived for the current system under consideration. These quantities are plotted in Fig.(\ref{comp_ExEy})(b).
In the system under consideration, the dynamical quantum optical lattice potential gets deeper with increase in the output photon number $|\alpha|^{2}$.  For the EBHM obtained in Eq.(\ref{Hbh}), the equivalent of $J$ defined for prototype BH model in (\ref{classicalBH}) are 
the tunnelling strengths $(\bar{E}_{x},\bar{E}_{y})$ and are defined in eqs. (\ref{EBHMpara1}) and (\ref{EBHMpara2}) in the Appendix \ref{expression}.
They become comparable to the corresponding $U_s$ near $\bar{\Delta}_c \sim$ $\bar{\Delta}_{p1,p2}= -3.37(-3.25)$ $MHz$ and beyond this point, $U_s$ starts dominating the tunnelling strengths, $(\bar{E}_{x},\bar{E}_{y})$, 
and eventually the tunnelling will be completely prohibited near $\bar{\Delta}_c =\bar{\Delta}_s$.

It may be pointed out that the relation (\ref{classicalBH}) is not rigorously valid for the EBHM defined in Eq.(\ref{Hbh}). Nevertheless, the good agreement between the values of the detuning parameter,
$\bar{\Delta}_{p}$ and $\bar{\Delta}_{p1,p2}$ evaluated in these two different ways does a consistency check on the EBHM parameters derived for the two pump cavity-atom system under consideration.
We, therefore, can conclude from the above discussion, that the phase coherence between two neighbouring wells in the dynamical quantum optical lattice potential is lost when the on-site interaction strength becomes comparable to the tunnelling strength \cite{Pethick, Stringari}, approximately at $\bar{\Delta}_c \sim \bar{\Delta}_{p1,p2}= -3.37(3.25)$ $MHz$, and the calculation from the HP approximation is less reliable beyond this point. The EBHM model on the other-hand can describe such phase-incoherent Mott-insulator phase in the system, even though we are not demonstrating the same in the current work.

In Fig.\ref{alphaEBHM}(c), we have plotted all the relevant Bose-Hubbard parameters as a function of $\bar{\Delta}_c$. These parameters naturally become non-zero for $\bar{\Delta}_c=\bar{\Delta}_{cr}$. As $\bar{\Delta}_c$ increases, $\alpha$ increases resulting in an increase in the depth of the dynamical optical lattice potential.
We already pointed out that this leads to increase in on-site energy $U_s$ and the simultaneous  decrease in the tunneling parameters $\tilde{E}_{x,y}$ as happened in  prototype Bose-Hubbard model in a classical optical lattice potential \cite{Zoller}.
\begin{figure*}
\includegraphics[width=2\columnwidth, height=1.5\columnwidth]{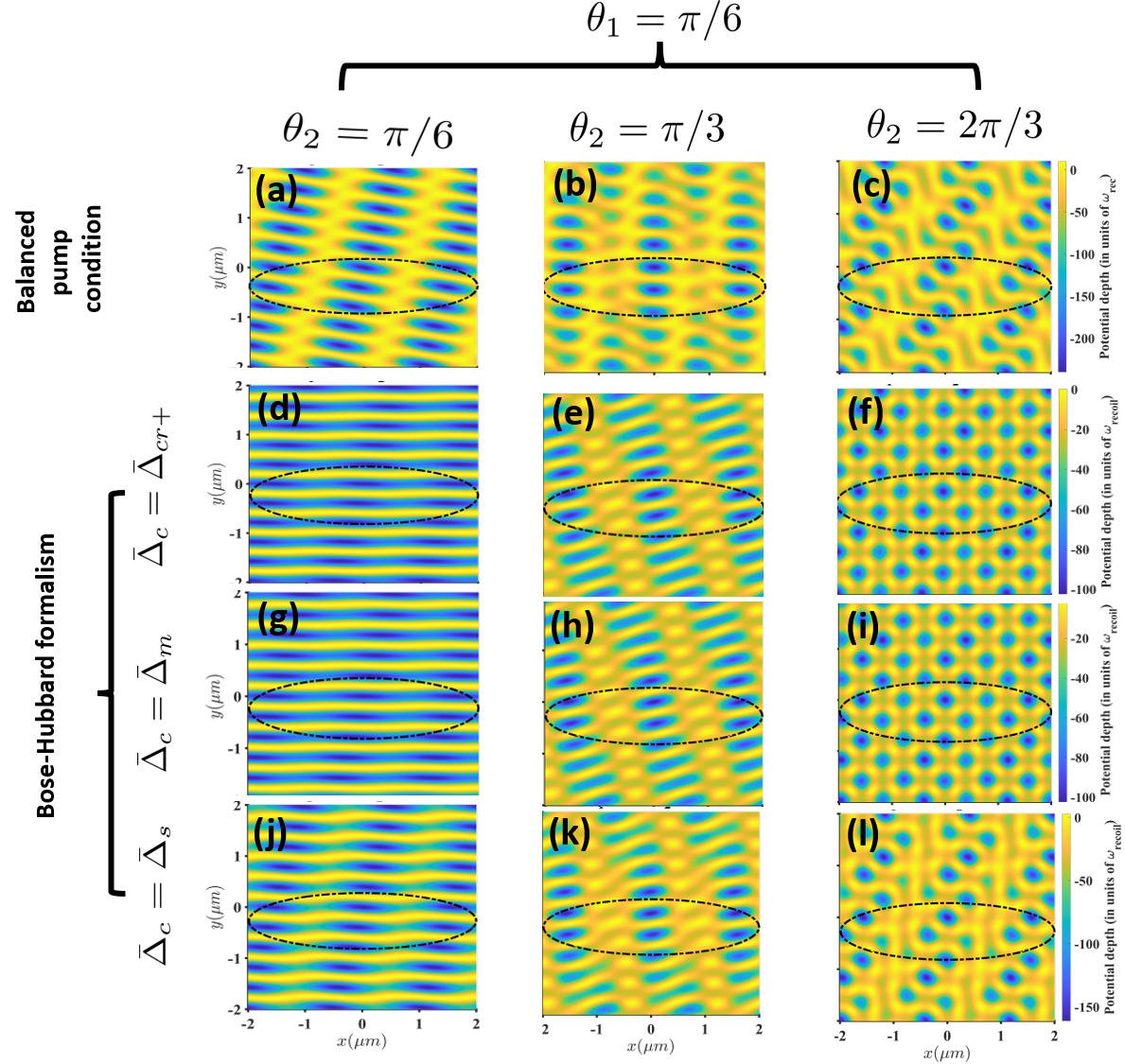}
 \caption{(\textit{color online}) \small (a)-(c) correspond to optical lattice potential as a function of $x$ and $y$ for balanced pump condition defined in section \ref{sec_QOL}. (d)-(l) corresponds to the evaluation of the dynamical optical lattice potential using EBHM and the method is detailed in section \ref{EBHM_QOL}.
 For these figures  the detuning and $\alpha$ increases in each column downward. In each row from left to right $\theta_{2}$ increases. $x$-axis remains same for all rows. Details are discussed in the text.
}\label{QOL}
\end{figure*}
The coefficient 
of the cavity mediated long range interaction term in the EBHM Hamiltonian (\ref{Hbh}), $U_{lx,ly, lxy}$ namely that appears before 
 $\hat{B}_{x}^2, \hat{B}_{y}^2$ and $\hat{B}_{x}\hat{B}_{y}$ respectively increases in absolute magnitude with increasing $\alpha$ implying stronger cavity mediated long-range interaction with more scattered photons, but these coefficients carry a negative sign as opposed to the positive sign of the onsite interaction 
 $U_{s}$. The presence of competing interaction terms in the EBHM Hamiltonian (\ref{Hbh}) with different signs explain the emergence of the lattice super-solid phase in such system. This EBHM based description, therefore, allows us to compare and contrast these systems with cavity mediated long-range interactions with
 a number of other continuum systems with competing long and short-range interactions that were simultaneously investigated for such supersolid phases, such as Rydberg atomic condensates \cite{Rydberg,Henkel}, dipolar bosonic quantum gases \cite{Lahaye, Lahaye1, Goral}. 

\subsection{Quantum optical lattice potential from Bose-Hubbard formalism}\label{EBHM_QOL} 
In the  Holstein-Primakoff approach discussed in Section \ref{HPA}, the atomic field operator $\hat{\Psi}(x,y)$ is expanded in five different modes in Eq. (\ref{atomic_op}). Whereas in the tight-binding approach that was used to derive the EBHM Hamiltonian (\ref{Hbh}), the atomic operator is expanded in terms of tightly bound Wannier orbitals 
in (\ref{eq3}), each of which is a superposition of Bloch waves of all allowed momentum values. Thus these two approaches use different Hilbert spaces for the expansion of the atomic field operators and a rigorous quantitative comparison between the results obtained from these two approaches is difficult to provide. 
In some limiting cases we can however make some conclusions 
based on the quantum optical lattice potentials evaluated using both these approaches and comparing them with the one evaluated under balanced pump condition. We shall do this with the help of Fig. (\ref{QOL}).
To evaluate the quantum optical lattice potential from the EBHM we note that since $J_x,J_y,\tilde{J}_{x1},\tilde{J}_{y1},\tilde{J}_{x2}$ and $\tilde{J}_{y2}$ are small as compared to $\tilde{J}_{01,02}$,  from Eq.(\ref{alpha_ss})
$\alpha = \braket{\hat{a}}$ can be well approximated by -
\bea
\alpha & = & \frac{\eta(\tilde{J}_{01}+\tilde{J}_{02})N}{\bar{\Delta}_c+i\kappa} \nonumber \\
\alpha+\alpha^* & = & \frac{2\bar{\Delta}_{c}\eta(\tilde{J}_{01}+\tilde{J}_{02})N}{\bar{\Delta}_c^2+\kappa^2} \label{QPBH} 
\eea 
In the first row, namely in Fig.(\ref{QOL})(a),(b),(c) we provide some representative plots of the optical lattice under the balanced pump condition, where $\alpha$ is not determined dynamically. 
 Ideally, the balance pump condition is obtained when $\alpha\rightarrow\infty$.
Under such balanced pump condition, we set $U_0\vert \alpha\vert^2=U_p$ and substitute this in the expression (\ref{Valpha}). 
 In the subsequent plots in Fig.(\ref{QOL}), namely, Fig. \ref{QOL}(d)-(l), we substitute $\vert\alpha\vert^2$ and $\alpha+\alpha^*$  calculated with Eq. \ref{QPBH} 
 in the expression (\ref{Valpha}) to obtain the quantum optical lattice potentials in EBHM. 
 
A comparison of Fig.(\ref{QOL}(a,b,c)) with the quantum optical lattice potential of Fig.(\ref{QOL}(j,k,l)) corresponding to eBHM, shows that the later potentials are approaching the balance pump condition as $\bar{\Delta}_{c}$ is increasing, namely when it approaches the cavity-pump resonance condition. 
The values of $\alpha$ obtained in the EBHM consider only the leading order terms and neglects the long range interactions but still it gives a significant agreement with the potential obtained using the balance pump condition. 
However, for other values of $\bar{\Delta}_{c}$ in the super-radiant regime, the profiles of the dynamical optical lattice potential obtained from HP approximation and plotted in Fig. \ref{OL_HP}(a)-(f) is considerably different from the one obtained under similar condition using BH approximation and plotted in Fig.(\ref{QOL}). 
It may be pointed out here that in the system considered, the super-radiance phenomenon is defined as the 
collective emission of light field by a group of $N$ atoms when they interact with a common light field \cite{Dicke} and emit light with an intensity proportional to $N^2$ \cite{Emary1}. The expression for $\alpha$ in (\ref{QPBH})  and the presence of $\alpha^{2}$ term in the expression of the dynamical optical potential (\ref{Valpha}) shows rigorously why the phases considered here beyond 
 the critical detuning are called super-radiant phases.

\section{Conclusions}
We have theoretically demonstrated a dimensional crossover in the self-organised lattice supersolid phases formed inside a linear cavity as a function of the relative angles between two classical pumps within a Holstein-Primakoff approach. We provided detailed classification of these self-organised phases by analysing their structures in co-ordinate and momentum spaces.
The corresponding quantum optical potential that is responsible for such super-radiant phases due to self organisation are plotted along with the corresponding atomic density and the relation between these plots are explained. In the later part of this work, we derived an effective Extended Bose-Hubbard Hamiltonian from the same microscopic Hamiltonian and with the help of the Bose-Hubbard parameters explain how cavity mediated long-range interaction is responsible for such super-solid phases. We also evaluated the dynamical optical lattice potential using the EBHM and compared it with the one obtained through the Holstein -Primakoff transformation and the one under balanced pump condition.  
Our proposal of observing such dimensional cross-over in a single set-up will hopefully augment further studies in this direction. The collective excitations accompanying such studies are another associated problem that can be looked into in future \cite{PSSG}. 
\section{ACKNOWLEDGMENTS}
This work is supported by a BRNS (DAE, Govt. of India)
Grant No. 21/07/2015-BRNS/35041 (DAE SRC Outstanding Investigator scheme). PS was also supported by an UGC ( Govt. of India) fellowship at the initial stage of this work.

\appendix

\setcounter{figure}{0} \renewcommand{\thefigure}{A.\arabic{figure}}

\section{Equations for mean field values $\Psi_{1\pm,2\pm}$ in Section(\ref{HPA})}\label{GSEQ} 
\begin{widetext} 
 \begin{subequations}
\begin{eqnarray}
\omega_{1+}\Psi_{1+}&-& \frac{2\lambda^2}{\bar{\Delta}_c}\Big(\Psi_{1+}\left(\Psi_{1+}+\Psi_{1-}\right)-\Psi_0^2\Big)\left(\Psi_{1+}+\Psi_{1-}\right)-\frac{2\lambda^2}{\bar{\Delta}_c}\Psi_{1+}\left(\Psi_{2+}+\Psi_{2-}\right)^2\nn\\
&-&\frac{2\lambda^2}{\bar{\Delta}_c}\Big(2\Psi_{1+}(\Psi_{1+}+\Psi_{1-})-\Psi^2_0\Big)(\Psi_{2+}+\Psi_{2-})=0\label{eqpsi1p}\\
\omega_{2+}\Psi_{2+}&-& \frac{2\lambda^2}{\bar{\Delta}_c}\Big(\Psi_{2+}\left(\Psi_{2+}+\Psi_{2-}\right)-\Psi_0^2\Big)\left(\Psi_{2+}+\Psi_{2-}\right)-\frac{2\lambda^2}{\bar{\Delta}_c}\Psi_{2+}\left(\Psi_{1+}+\Psi_{1-}\right)^2\nn\\
&-&\frac{2\lambda^2}{\bar{\Delta}_c}\Big(2\Psi_{2+}(\Psi_{2+}+\Psi_{2-})-\Psi^2_0\Big)(\Psi_{1+}+\Psi_{1-})=0\label{eqpsi2p}\\
\omega_{1-}\Psi_{1-}&-& \frac{2\lambda^2}{\bar{\Delta}_c}\Big(\Psi_{1-}\left(\Psi_{1+}+\Psi_{1-}\right)-\Psi_0^2\Big)\left(\Psi_{1+}+\Psi_{1-}\right)-\frac{2\lambda^2}{\bar{\Delta}_c}\Psi_{1-}\left(\Psi_{2+}+\Psi_{2-}\right)^2\nn\\
&-&\frac{2\lambda^2}{\bar{\Delta}_c}\Big(2\Psi_{1-}(\Psi_{1+}+\Psi_{1-})-\Psi^2_0\Big)(\Psi_{2+}+\Psi_{2-})=0\label{eqpsi1m}\\
\omega_{2-}\Psi_{2-}&-& \frac{2\lambda^2}{\bar{\Delta}_c}\Big(\Psi_{2-}\left(\Psi_{2+}+\Psi_{2-}\right)-\Psi_0^2\Big)\left(\Psi_{2+}+\Psi_{2-}\right)-\frac{2\lambda^2}{\bar{\Delta}_c}\Psi_{2-}\left(\Psi_{1+}+\Psi_{1-}\right)^2\nn\\
&-&\frac{2\lambda^2}{\bar{\Delta}_c}\Big(2\Psi_{2-}(\Psi_{2+}+\Psi_{2-})-\Psi^2_0\Big)(\Psi_{1+}+\Psi_{1-})=0\label{eqpsi2m}
\end{eqnarray}
\end{subequations}
\end{widetext}

\section{Derivation for critical detuning in Section(\ref{HPA})}\label{CRD}
In this section we provide a derivation for Eq.(\ref{crdelta}) and Eq.(\ref{crdeltakappa}). We first find out the Hessian matrix for the Hamiltonian in Eq.(\ref{Halpha}) which is given as \cite{Emary}.

\bea
\begin{bmatrix}
\frac{\partial^2 h_{m=0}^{(0)}}{\partial \Psi_{1+}^2} &\frac{\partial^2 h_{m=0}^{(0)}}{\partial \Psi_{1+}\Psi_{1-}}&\frac{\partial^2 h_{m=0}^{(0)}}{\partial \Psi_{1+}\Psi_{2+}}&
\frac{\partial^2 h_{m=0}^{(0)}}{\partial \Psi_{1+}\Psi_{2-}}\\
\frac{\partial^2 h_{m=0}^{(0)}}{\partial \Psi_{1-}\Psi_{1+}}&\frac{\partial^2 h_{m=0}^{(0)}}{\partial \Psi_{1-}^2}&
\frac{\partial^2 h_{m=0}^{(0)}}{\partial \Psi_{1-}\Psi_{2+}}&
\frac{\partial^2 h_{m=0}^{(0)}}{\partial \Psi_{1-}\Psi_{2-}}\\
\frac{\partial^2 h_{m=0}^{(0)}}{\partial \Psi_{2+}\Psi_{1+}}&
\frac{\partial^2 h_{m=0}^{(0)}}{\partial \Psi_{2+}\Psi_{1-}}&\frac{\partial^2 h_{m=0}^{(0)}}{\partial \Psi_{2+}^2}&
\frac{\partial^2 h_{m=0}^{(0)}}{\partial \Psi_{2+}\Psi_{2-}}\\
\frac{\partial^2 h_{m=0}^{(0)}}{\partial \Psi_{2-}\Psi_{1+}}&
\frac{\partial^2 h_{m=0}^{(0)}}{\partial \Psi_{2-}\Psi_{1-}}&
\frac{\partial^2 h_{m=0}^{(0)}}{\partial \Psi_{2-}\Psi_{2+}}&\frac{\partial^2 h_{m=0}^{(0)}}{\partial \Psi_{2-}^2}\nn
\end{bmatrix}
\eea
In normal phase $\Psi_{1+}=\Psi_{2+}=\Psi_{1-}=\Psi_{2-}=0$, therefore, $\Psi_{0}=1$. The Hessian takes the following form -
\bea
\begin{bmatrix}
2\hbar \omega_{1+}+\frac{8\hbar\lambda^2}{\bar{\Delta}_c}&\frac{8\hbar\lambda^2}{\bar{\Delta}_c}&\frac{8\hbar\lambda^2}{\Delta_c}&\frac{8\hbar\lambda^2}{\bar{\Delta}_c}\\
\frac{8\hbar\lambda^2}{\bar{\Delta}_c}&2\hbar \omega_{1-}+\frac{8\hbar\lambda^2}{\bar{\Delta}_c}&\frac{8\hbar\lambda^2}{\bar{\Delta}_c}&\frac{8\hbar\lambda^2}{\bar{\Delta}_c}\\
\frac{8\hbar\lambda^2}{\bar{\Delta}_c}&\frac{8\hbar\lambda^2}{\bar{\Delta}_c}&2\hbar \omega_{2+}+\frac{8\hbar\lambda^2}{\bar{\Delta}_c}&\frac{8\hbar\lambda^2}{\bar{\Delta}_c}\\
\frac{8\hbar\lambda^2}{\bar{\Delta}_c}&\frac{8\hbar\lambda^2}{\bar{\Delta}_c}&\frac{8\hbar\lambda^2}{\bar{\Delta}_c}&2\hbar \omega_{2-}+\frac{8\hbar\lambda^2}{\bar{\Delta}_c}\nn
\end{bmatrix}
\eea
At critical detuning, $\bar{\Delta}_{cr}$, $\Psi_{1+}=\Psi_{2+}=\Psi_{1-}=\Psi_{2-}\neq 0$, therefore, $\Psi_{0}\neq 1$. Therefore, at this point, the determinant of the Hessian gives us the critical detuning - 
\bea
\bar{\Delta}_{cr}=-\frac{4}{\bar{\omega}_{1}}\lambda^2 -\frac{4}{\bar{\omega}_{2}}\lambda^2 
\eea
where 
\bea
\bar{\omega}_{1+}&=&\frac{\omega_{1+}\omega_{1-}}{\omega_{1+}+\omega_{1-}}\nn\\
\bar{\omega}_{2+}&=&\frac{\omega_{2+}\omega_{2-}}{\omega_{2+}+\omega_{2-}}\nn
\eea
In presence of atom-atom interactions, $g_{2D}$ and cavity decay rate, $\kappa$, the critical detuning gets modified and is given as -
\begin{eqnarray}
\bar{\Delta}_{cr} = -\frac{2\lambda^2 }{\omega_{10}}-\sqrt{\frac{-4\lambda^4 }{\omega_{10}^2}-\kappa^2}-\frac{2\lambda^2 }{\omega_{20}}-\sqrt{\frac{-4\lambda^4 }{\omega_{20}^2}-\kappa^2}\nn
\end{eqnarray}

\section{Derivation of the maxima and minima points in the atomic density in Section (\ref{class})}\label{MAXMIN} 
\begin{widetext}
In this section we shall describe the analytical technique of obtaining the maxima and minima points of the atomic density for special combination of angles $\theta_{1}$ and $\theta_{2}$.
In general such maxima and minima has to be obtained numerically. 
\begin{eqnarray}
&&\frac{\partial \vert\Psi(x,y)\vert^2}{\partial x}=0\nn\\
&&\Rightarrow\Psi_{1+} \sin(kx \sin\theta_1+k y (1+\cos\theta_1))\sin\theta_1\nn\\
&&+\Psi_{2+}\sin(k x (\sin\theta_1-\sin(\theta_2-\theta_1)) + k y  (\cos\theta_1+\cos(\theta_2-\theta_1)))(\sin\theta_1-\sin(\theta_2-\theta_1))\nn\\
&&+\Psi_{1-} \sin(kx \sin\theta_1+k y (-1+\cos\theta_1))\sin\theta_1\nn\\
&&+\Psi_{2-}\sin(k x (\sin\theta_1+\sin(\theta_2-\theta_1)) + k y  (\cos\theta_1-\cos(\theta_2-\theta_1)))(\sin\theta_1+\sin(\theta_2-\theta_1))=0\label{der_x}
\end{eqnarray}
\begin{eqnarray}
&&\frac{\partial \vert\Psi(x,y)\vert^2}{\partial y}=0\nn\\
&&\Rightarrow\Psi_{1+} \sin(kx \sin\theta_1+k y (1+\cos\theta_1))(1+\cos\theta_1)\nn\\
&&+\Psi_{2+}\sin(k x (\sin\theta_1-\sin(\theta_2-\theta_1)) + k y  (\cos\theta_1+\cos(\theta_2-\theta_1)))(\cos\theta_1+\cos(\theta_2-\theta_1))\nn\\
&&+\Psi_{1-} \sin(kx \sin\theta_1 + k y (-1+\cos\theta_1))(-1+\cos\theta_1)\nn\\
&&+\Psi_{2-}\sin(k x (\sin\theta_1+\sin(\theta_2-\theta_1)) + k y  (\cos\theta_1-\cos(\theta_2-\theta_1)))(\cos\theta_1-\cos(\theta_2-\theta_1))=0\label{der_y}
\end{eqnarray}
\end{widetext}
For $\theta_2-\theta_1=0$, the corresponding maxima and minima points in the atomic density plot can be obtained analytically. To demonstrate that we substitute 
$\theta_{1} = \theta_{2}$ in Eqs. \ref{der_x} and \ref{der_y} to get 
\begin{widetext}
\begin{eqnarray}
\partial_x \vert\Psi(x,y)\vert^2 &=& \Psi_{1+} \sin(kx \sin\theta_1+k y +k y\cos\theta_1)\sin\theta_1\nn\\
&&+\Psi_{2+}\sin(k x \sin\theta_1 + k y  +k y \cos\theta_1)\sin\theta_1\nn\\
&&+\Psi_{1-} \sin(kx \sin\theta_1 -k y +k y\cos\theta_1)\sin\theta_1\nn\\
&&+\Psi_{2-}\sin(k x \sin\theta_1 + k y  \cos\theta_1-ky)\sin\theta_1\nn\\
&\Rightarrow& \partial_x \vert\Psi(x,y)\vert^2 =0
\end{eqnarray}
\end{widetext}
\begin{widetext}
\begin{eqnarray}
\partial_y\vert\Psi(x,y)\vert^2 &=& \Psi_{1+} \sin(kx \sin\theta_1 +k y +k y\cos\theta_1)(1+\cos\theta_1)\nn\\
&&+\Psi_{2+}\sin(k x \sin\theta_1 + k y  +k y \cos\theta_1)(1+\cos\theta_1)\nn\\
&&+\Psi_{1-} \sin(kx \sin\theta_1-k y +k y\cos\theta_1)(-1+\cos\theta_1)\nn\\
&&+\Psi_{2-}\sin(k x \sin\theta_1 + k y  \cos\theta_1-ky)(-1+\cos\theta_1)\nn\\
&\Rightarrow& \partial_y \vert\Psi(x,y)\vert^2 =0
\end{eqnarray}
\end{widetext}
Since $\Psi_{1\pm}$ and $\Psi_{2\pm}$ are independent non-zero momentum component of the superfluid order parameter, the solutions of the 
above equation can be obtained from 
\begin{eqnarray}
kx \sin\theta_1 +ky+ ky\cos\theta_1 &=& n\pi\nn, n \in \mathcal{I}\\
kx \sin\theta_1 -ky+ ky\cos\theta_1 &=&  m\pi\nn n \in \mathcal{I} 
\end{eqnarray}
whose solution gives us the co-ordinates of $x$ and $y$ where the densities are extremum, namely 
\bea
x &=& \frac{\lambda_p((n+m)-(n-m)\cos\theta_1)}{4\sin\theta_1}\nn\\
y &=& \frac{(n-m)\lambda_p}{4}\nn
\eea
 To find out the maxima and minima points we need to evaluate the $F_{xx}$ and $F_{yy}$ at these points and obtain that $F_{xx}<0$ and $F_{yy}<0$ when $n$ and $m$ are $\text{even}$ integers and $F_{xx}>0$ and $F_{yy}>0$ when $n$ and $m$ are $\text{odd}$ integers.

\section{Relation between our model and other BH models}\label{Rel_EBHM}
A comparison of our BH model in Eq.(\ref{Hbh}) with the models considered in \cite{Maschler, Landig, Dogra, Ritsch}, also reveals that the long-range interaction terms of our model are proportional to $\hat{B}_{x,y}^2$ which is different from the models in \cite{Maschler, Landig, Dogra, Ritsch}, where the global-range interactions favour particle imbalance between odd and even sites. The difference in the origin of the infinite range interactions in the two models is a consequence of the lattice geometry. The optical lattice potential in \cite{Maschler, Landig, Dogra, Ritsch} has equal depths along the $x-$ and the $y-$ directions which gives rise to a square lattice while in our case we have different lattice depths along the $x-$ and the $y-$ direction and we get a distorted square lattice for $\bar{\Delta}_c > \bar{\Delta}_{m}$. 
\section{Expressions for Bose-Hubbard model in Section(\ref{BHM})}\label{expression}
The hopping amplitudes along the $x$ and $y$ direction and on-site energy are given by
\begin{widetext}
\bea 
E_{x} & = & \frac{1}{2}\int\int dxdy\ w_{p,q}^{*}(x,y)\left( -\frac{\hbar^{2}}{2M_a}\nabla^2 + \hbar U_p \cos^2(\bs{k}_1\cdot\bs{r}) +\hbar U_p \cos^2(\bs{k}_2\cdot\textbf{r}) + \hbar U_p \cos(\bs{k}_1\cdot\bs{r})\cos(\bs{k}_2\cdot\bs{r})\right)w_{p+1,q}(x,y) \nn  \\
E_{y} &=& \frac{1}{2}\int\int dxdy\ w_{p,q}^{*}(x,y)\left( -\frac{\hbar^{2}}{2M_a}\nabla^2 + \hbar U_p \cos^2(\bs{k}_1\cdot\bs{r}) +\hbar U_p \cos^2(\bs{k}_2\cdot\bs{r}) + \hbar U_p \cos(\bs{k}_1\cdot\bs{r})\cos(\bs{k}_2\cdot\bs{r})\right)w_{p,q+1}(x,y) \nn \\
E_{0}&=&\int\int dxdy\ w_{p,q}^{*}(x,y)\left( -\frac{\hbar^{2}}{2M_a}\nabla^2 + \hbar U_p \cos^2(\bs{k}_1\cdot\textbf{r}) +\hbar U_p \cos^2(\bs{k}_2\cdot\bs{r}) + \hbar U_p \cos(\bs{k}_1\cdot\bs{r})\cos(\bs{k}_2\cdot\bs{r})\right)w_{p,q}(x,y) \nn
\eea 
\end{widetext}
The hopping and onsite interactions due to the photons scattered by the atoms are given as 
\begin{widetext}
\bea 
J_{x} &=& \frac{1}{2}\int\int dxdy\ w_{p,q}^{*}(x,y)\ \cos^{2}(\bs{k}_{c}\cdot\bs{r}) w_{p+1,q}(x,y)\nonumber \\
J_{y}&=& \frac{1}{2}\int\int dxdy\ w_{p,q}^{*}(x,y)\ \cos^{2}(\bs{k}_{c}\cdot\bs{r}) w_{p,q+1}(x,y)\nonumber \\
J_{0} &=& \int\int dx dy\ \vert w_{p,q}(x,y)\vert^2\ \cos^{2}(\bs{k}_{c}\cdot\bs{r})\nonumber \\
\tilde{J}_{01} &=& \int\int dxdy\ \vert w_{p,q}(x,y)\vert^2\ \cos(\bs{k}_{c}\cdot\bs{r})\cos(\bs{k}_{1}\cdot\bs{r}) \nonumber\\
\tilde{J}_{02} &=& \int\int dxdy\ \vert w_{p,q}(x,y)\vert^2\ \cos(\bs{k}_{c}\cdot\bs{r})\cos(\bs{k}_{2}\cdot\bs{r}) \nonumber\\
\tilde{J}_{x1} &=& \frac{1}{2}\int\int dxdy\ w^*_{p,q}(x,y)\ \cos(\bs{k}_{c}\cdot\bs{r})\cos(\bs{k}_{1}\cdot\bs{r} )w_{p+1,q}(x,y) \nonumber\\
\tilde{J}_{x2} &=& \frac{1}{2}\int\int dxdy\ w^*_{p,q}(x,y)\ \cos(\bs{k}_{c}\cdot\bs{r})\cos(\textbf{k}_{2}\cdot\bs{r} )w_{p+1,q}(x,y) \nonumber\\
\tilde{J}_{y1} &=& \frac{1}{2}\int\int dxdy\ w^*_{p,q}(x,y)\ \cos(\bs{k}_{c}\cdot\bs{r})\cos(\textbf{k}_{1}\cdot\bs{r} )w_{p,q+1}(x,y) \nonumber\\
\tilde{J}_{y2} &=& \frac{1}{2}\int\int dxdy\ w^*_{p,q}(x,y)\ \cos(\bs{k}_{c}\cdot\bs{r})\cos(\bs{k}_{2}\cdot\bs{r} )w_{p,q+1}(x,y) \nonumber
\eea
\end{widetext}
The detailed expression of the Bose-Hubbard parameters that appear in the Hamiltonian (\ref{Hbh})  are listed below 
\bea
\tilde{E}_x &=& E_x+\frac{2\hbar \bar{\Delta}_c\eta^2(\tilde{J}_{x1}+\tilde{J}_{x2})(\tilde{J}_{01}+\tilde{J}_{02})N}{\bar{\Delta}_c^2 +\kappa^2} \nn \\
&+& \frac{2\hbar\eta^2 N^2U_0J_x(\bar{\Delta}_c^2-\kappa^2)(\tilde{J}_{01}+\tilde{J}_{02})^2}{(\bar{\Delta}_c^2 +\kappa^2)^2}  \label{EBHMpara1} \\ 
\tilde{E}_y &=& E_y+\frac{2\hbar \bar{\Delta}_c\eta^2(\tilde{J}_{y1}+\tilde{J}_{y2})(\tilde{J}_{01}+\tilde{J}_{02})N}{\bar{\Delta}_c^2 +\kappa^2} \nn \\
&+&\frac{2\hbar\eta^2 N^2U_0J_y(\bar{\Delta}_c^2-\kappa^2)(\tilde{J}_{01}+\tilde{J}_{02})^2}{(\bar{\Delta}_c^2 +\kappa^2)^2} \label{EBHMpara2} 
\eea
\bea
U_{lx} &=& \frac{\hbar \bar{\Delta}_c\eta^2 (\tilde{J}_{x1}+\tilde{J}_{x2})^2}{\bar{\Delta}^2_c +\kappa^2}\nn\\
&+& \frac{2\hbar\eta^2 NU_0J_x(\bar{\Delta}_c^2-2\kappa^2)(\tilde{J}_{01}+\tilde{J}_{02})(\tilde{J}_{x1}+\tilde{J}_{x2})}{(\bar{\Delta}_c^2 +\kappa^2)^2}\nn\\
U_{ly} &=& \frac{\hbar \bar{\Delta}_c\eta^2 (\tilde{J}_{y1}+\tilde{J}_{y2})^2}{\bar{\Delta}^2_c +\kappa^2}\nn\\
&+& \frac{2\hbar\eta^2 NU_0J_y(\bar{\Delta}_c^2-2\kappa^2)(\tilde{J}_{01}+\tilde{J}_{02})(\tilde{J}_{y1}+\tilde{J}_{y2})}{(\bar{\Delta}_c^2 +\kappa^2)^2}\nn
\eea

\end{document}